\documentclass[sigconf]{acmart}

\AtBeginDocument{%
  \providecommand\BibTeX{{%
    \normalfont B\kern-0.5em{\scshape i\kern-0.25em b}\kern-0.8em\TeX}}}

\copyrightyear{2022}
\acmYear{2022}
\setcopyright{rightsretained}
\acmConference[SIGMOD '22] {Proceedings of the 2022 International Conference on Management of Data}{June 12--17, 2022}{Philadelphia, PA, USA.}
\acmBooktitle{Proceedings of the 2022 International Conference on Management of Data (SIGMOD '22), June 12--17, 2022, Philadelphia, PA, USA}
\acmPrice{}
\acmISBN{978-1-4503-9249-5/22/06}
\acmDOI{10.1145/3514221.3517885}

\usepackage{graphicx}
\usepackage{balance}  %
\pdfoutput=1

\usepackage{enumitem}
\usepackage{amstext}
\usepackage{alltt}
\usepackage{epstopdf}
\usepackage{xspace,colortbl}
\usepackage{multicol}
\usepackage{algorithm}
\usepackage{algorithmicx}
\usepackage[noend]{algpseudocode}
\usepackage{url}
\usepackage{hyperref}

\usepackage[font={small}]{caption}

\usepackage{bbold}
\usepackage{dsfont}
\usepackage{bm}
\usepackage{csquotes}
\usepackage{courier}

\usepackage{float}
\usepackage{listings}
\newfloat{lstfloat}{htbp}{lop}
\floatname{lstfloat}{Listing}
\definecolor{dkgreen}{rgb}{0,0.6,0}
\definecolor{ltgray}{rgb}{0.5,0.5,0.5}

\usepackage{adjustbox}

\definecolor{light-gray}{gray}{0.95}
\definecolor{mid-gray}{gray}{0.85}
\definecolor{green}{RGB}{0,176,80}
\definecolor{darkred}{rgb}{0.7,0.25,0.25}
\definecolor{darkgreen}{rgb}{0.15,0.55,0.15}
\definecolor{darkblue}{rgb}{0.1,0.1,0.5}
\definecolor{orange}{RGB}{237,125,49}
\definecolor{blue}{RGB}{68,114,196}
\definecolor{pop}{RGB}{0,21,245}
\definecolor{myblue}{RGB}{59,106,144}
\definecolor{realblue}{RGB}{0,0,255}
\definecolor{airforceblue}{rgb}{0.36, 0.54, 0.66}
\definecolor{LightCyan}{rgb}{0.88,1,1}

\definecolor{newblue}{RGB}{162,212,236}

\usepackage{xcolor}
\usepackage{soul}
\sethlcolor{LightCyan}

\usepackage{verbatim}
\usepackage{subcaption}
\newsavebox{\imagebox}
\usepackage{booktabs}
\usepackage{tabularx}

\DeclareCaptionType{Example}

\usepackage{pifont}%

\usepackage{bbm}

\newcommand{\secref}[1]{\S\ref{#1}}
\newcommand{\pg}{{PostgreSQL}\xspace}
\newcommand{\dbms}{{CommDB}\xspace}

\let\oldsim\sim
\renewcommand{\sim}{{\oldsim}}

\newcommand{\etal}{\emph{et~al.}\xspace}

{\endlist}

\makeatletter
\newcommand{\distas}[1]{\mathbin{\overset{#1}{\kern\z@\sim}}}%
\newsavebox{\mybox}\newsavebox{\mysim}
\newcommand{\distras}[1]{%
  \savebox{\mybox}{\hbox{\kern3pt$\scriptstyle#1$\kern3pt}}%
  \savebox{\mysim}{\hbox{$\sim$}}%
  \mathbin{\overset{#1}{\kern\z@\resizebox{\wd\mybox}{\ht\mysim}{$\sim$}}}%
}
\makeatother

\newcommand{\vsim}{$V_\text{sim}$\xspace}
\newcommand{\vreal}{$V_\text{real}$\xspace}

\newcommand{\cout}{$C_{out}$\xspace}
\begin{document}

\newcommand{\sys}{Balsa\xspace}

\title{\sys: Learning a Query Optimizer Without Expert Demonstrations}
\fancyhead{}

\author{Zongheng Yang}
\affiliation{%
  \institution{UC Berkeley}
  \city{Berkeley}
  \state{CA}
  \country{USA}
}
\author{Wei-Lin Chiang}
\authornote{Equal contribution.}
\affiliation{%
  \institution{UC Berkeley}
  \city{Berkeley}
  \state{CA}
  \country{USA}
}
\author{Sifei Luan}
\authornotemark[1]
\affiliation{%
  \institution{UC Berkeley}
  \city{Berkeley}
  \state{CA}
  \country{USA}
}
\author{Gautam Mittal}
\affiliation{%
  \institution{UC Berkeley}
  \city{Berkeley}
  \state{CA}
  \country{USA}
}
\author{Michael Luo}
\affiliation{%
  \institution{UC Berkeley}
  \city{Berkeley}
  \state{CA}
  \country{USA}
}
\author{Ion Stoica}
\affiliation{%
  \institution{UC Berkeley}
  \city{Berkeley}
  \state{CA}
  \country{USA}
}

\begin{abstract}

Query optimizers are a performance-critical component in every database system.
Due to their complexity, optimizers take experts months to write and years to refine.
In this work, we demonstrate for the first time that {learning to optimize queries without learning from an expert optimizer}
is both possible and efficient.
We present \sys, a query optimizer {built by} deep reinforcement learning.
{\sys first learns basic knowledge from a simple, environment-agnostic simulator, followed by safe learning in real execution.}
{On the Join Order Benchmark, \sys matches the performance of two expert query optimizers, both open-source and commercial, with two hours of learning, and outperforms them by up to $2.8\times$ in workload runtime after a few more hours}.
\sys thus opens the possibility of automatically learning to optimize in future compute environments where expert-designed optimizers do not exist.

\end{abstract}



\maketitle

\section{Introduction}
\label{sec:intro}

Query optimizers are a performance-critical component in every database and query engine, translating declarative queries into efficient execution plans. These optimizers must navigate a vast search space of candidate plans for each query, scoring each plan with sufficient accuracy by leveraging statistics about the data.

As a result of this complexity, optimizers are costly to develop.
Human experts may spend months to write a first version and years to refine it. For example, PostgreSQL, one of the most widely used databases in the world, has seen a continuous stream of changes to its optimizer more than 20 years after it was released~\cite{pgoptimizer}. Due to the high development costs, some relational systems settle for heuristic-based optimizations and postpone building full-fledged cost-based optimizers.
As examples, Spark SQL was introduced in 2014 but only added a cost-based optimizer (CBO) in 2017, while CockroachDB shipped the first version of its CBO in v2.1 after ``9 months of intense effort''~\cite{cockroach_blog}.

Instead of having human experts spend years developing a state-of-the-art optimizer, in this paper we ask whether it is possible to use machine learning to
\emph{{learn to optimize queries without learning from an existing expert optimizer.}}
We answer this question affirmatively by designing and implementing \sys, a learned query optimizer that can match or even exceed the performance of {expert-built query optimizers (both open-source and commercial)}. %

\sys leverages deep reinforcement learning (RL), which has been successfully employed to learn complex skills~\cite{openai_cube} and exceed human experts at playing games~\cite{Silver2016,Silver2017,alphastar}.
RL consists of an \emph{agent} that learns to solve a task by repeatedly interacting with an \emph{environment}.
The agent observes the environment's \emph{state} and takes an \emph{action} to maximize a \emph{reward}.
If the actions
lead to improved rewards,
they are \emph{reinforced}, i.e., the agent is updated to make these actions more likely in the future.
For a learned optimizer agent, such as \sys, the environment is the database;
a state is a partial plan for a query;
an action is to add operators to the partial plan,
and the reward for a complete plan is its execution latency (negated).
Using this feedback loop, \sys learns by trial and error to become increasingly
better at generating query execution plans.

In fact, the promise of RL for query optimization has been shown by several recent projects{~\cite{dq,neo,marcus2020bao}}.
However, these methods assume the availability of a mature query optimizer to learn from.
In contrast, {\sys does not learn from such an expert optimizer.}
To our knowledge, \sys demonstrates for the first time that \emph{learning to optimize queries {without learning from an expert optimizer} is both possible and efficient}.
This can have a far reaching impact, as it paves the road towards automatically learning to optimize in new data systems~\cite{modin,materialize_blog} where a mature optimizer does not exist.

A unique challenge in learning {to optimize queries without an expert optimizer's guidance} %
is that most execution plans for a query are slow---sometimes orders of magnitude more expensive than the optimal plan~\cite{leis2015good,leis2018query}.
At the beginning of the learning process, the agent has no prior knowledge, so the probability of selecting such {disastrous} plans is high,
{which may prevent any progress.} %
This is a unique characteristic of query optimization that is not shared by other successful RL applications such as games.
Indeed, with most games (e.g., AlphaGo~\cite{Silver2016}, MuZero~\cite{muzero}), a ``bad'' action typically leads to a game ending quicker.
As a result, bad actions do not {hinder learning in those environments}.

{
To avoid disastrous plans,
\sys employs simulation-to-reality learning~\cite{tobin2017domain}.
In the ``simulation'' phase, \sys quickly learns from a simulator how to avoid disastrous plans without executing queries, while in the ``reality'' phase it learns from real executions to produce high-performance plans.
The simulator gives cost feedback to the agent by using a basic, logical-only cost model with a cardinality estimator.
For convenience, we use \pg's cardinality estimator, a simple histogram-based method~\cite{leis2015good}.
We pick an existing estimator since, unlike an optimizer, a cardinality estimator is agnostic to the execution environment, so the same estimator can be used for any environment. 
(In our evaluation, we use \pg's estimates for another commercial engine.)
Moreover, the estimator needs not be high-quality for effective simulation.
In fact, \pg's estimates can exhibit orders of magnitude errors~\cite{leis2015good}, and we find that even injecting noises to these estimates does not impact \sys's performance (\secref{sec:discussions}). This is because \sys only uses the simulation to learn to avoid disastrous plans, not to reach expert-level performance. Therefore, basic cost models and estimates suffice.
}

{Next, to vastly improve over the imperfect knowledge acquired from the simulator, \sys learns in the real environment by actually executing queries.}
While the simulation knowledge enables the agent to avoid the worst plans, it can still stumble onto %
bad plans, causing unpredictable stalls in the learning process.
\sys addresses this challenge by using \emph{timeouts}.
A query's timeout is set to its best latency so far during learning.
If a plan times out, we assign it a predefined
low reward (as we do not know its true reward).
If the plan finishes, we tighten the timeout for future iterations. Thus, timeouts bound each learning iteration's runtime, ensuring \emph{safe execution}
that eliminates unpredictable stalls.

Finally, {an RL agent must balance exploiting past experiences with exploring new ones to escape local minima.
The classic solution is random exploration, i.e., occasionally pick a random plan.
Unfortunately, this standard strategy is ineffective, since random plans in the search space are likely to be highly expensive.}
Instead, \sys explores from a set of \emph{probably good} plans.
During exploration, \sys generates \emph{several} best predicted plans (instead of the best), then picks the best unseen one out of them.
This \emph{safe exploration} approach {improves \sys's plan coverage and performance.}

Given a target dataset, \sys is trained by repeatedly optimizing a set of sample queries by trial and error.
After training, we test its \emph{generalization} performance on a new set of {unseen} queries for the same dataset.
We find that all three components of \sys---simulation learning, safe execution, safe exploration---boost
its
generalization.
They expose \sys to a higher quantity and variety of plans, thereby enabling it to optimize new queries more robustly---a trait we believe is essential for the practical deployment of learned optimizers.
{We further propose using \emph{diversified experiences} to enhance generalization (\secref{sec:generalization}).}
We study \sys's generalization in depth in our evaluation (\secref{sec:eval:perf}, \secref{sec:eval-generalization}), and
find that
it achieves {better performance than two expert optimizers on unseen queries}.

We call our approach
``\textbf{B}ootstrap, S\textbf{a}fe\textbf{l}y Execute, \textbf{Sa}fely Explore'', hence \sys\footnote{Balsa wood is famous for its light weight.} for short. %
To
our knowledge, \sys is the first learned optimizer that does not rely on plans {(\emph{demonstrations})} generated by an existing expert optimizer.
On the Join Order Benchmark~\cite{leis2015good},
a complex workload designed to stress test optimizers,
\sys matches the performance of {two expert optimizers with two hours of training, and outperforms them by 2.1--2.8$\times$ after a few more hours.}

In summary, we make the following contributions:%
\begin{itemize}[leftmargin=*]
\item We {introduce \sys, a learned query optimizer that does not learn from an existing, expert optimizer.}
\item We design a simple approach for learning a query optimizer {without expert demonstrations}: bootstrapping from simulation (\secref{sec:sim_learning}), safe execution (\secref{sec:real_learning}), and safely exploring the plan space (\secref{sec:exploration}).
\item {We propose \emph{diversified experiences}, a novel method to further enhance training and generalization performance (\secref{sec:generalization}), including generalizing to unseen queries with highly distinct join templates.}
\item {\sys can outperform both an open-source (\pg) and a commercial query optimizer, after a few hours of training (\secref{sec:eval}).}
\item {We show that, despite not learning from an expert optimizer, \sys outperforms the prior state-of-the-art technique that does.}
\end{itemize}
\sys is open sourced at \url{https://github.com/balsa-project/balsa}.

\subsection{{Differences from Prior Work}}

{
To highlight \sys's contributions, we briefly compare with the most related work and defer a complete discussion to \secref{sec:related}.

\emph{DQ}~\cite{dq} learns from an expert optimizer's \emph{cost model}. As such, its performance is bounded by the quality of the cost model, which can be inaccurate.
\emph{Neo}~\cite{neo} takes an opposite approach by learning from an expert optimizer's \emph{plans} and real executions. While this is more accurate than using just a cost model, it is also more expensive.
Importantly, these solutions assume either an \emph{expert} cost model or an \emph{expert} optimizer to bootstrap from.

In contrast, \sys requires neither an expert cost model (as in DQ) nor an expert optimizer
(as in Neo) to learn from.
\sys removes these fundamental assumptions by bootstrapping from a minimal, logical-only cost model, followed by safe learning in real execution.
For the cost model, \sys needs a basic cardinality estimator (\secref{sec:sim-discussion}). %
We find inaccurate estimates can still lead to successful simulation, and most
of \sys's knowledge is learned after simulation (\secref{sec:discussions}).

In summary, this paper tackles the \emph{new problem} of learning to optimize when an expert optimizer does not exist.
(We
discuss in \secref{sec:discussions} how \sys can \emph{better leverage} an expert, if available, than prior work.)
To solve this problem, we develop or apply techniques new to the domain of learned optimizers. %
These include sim-to-real (\secref{sec:overview}), safe execution (\secref{sec:timeout}), safe exploration (\secref{sec:exploration}), on-policy learning (\secref{sec:rl-algo}), and enhancing generalization with diversified experiences (\secref{sec:generalization}).

}

\section{\sys Overview}
\label{sec:overview}

\begin{figure}[tp]
\centering
\includegraphics[width=\columnwidth]{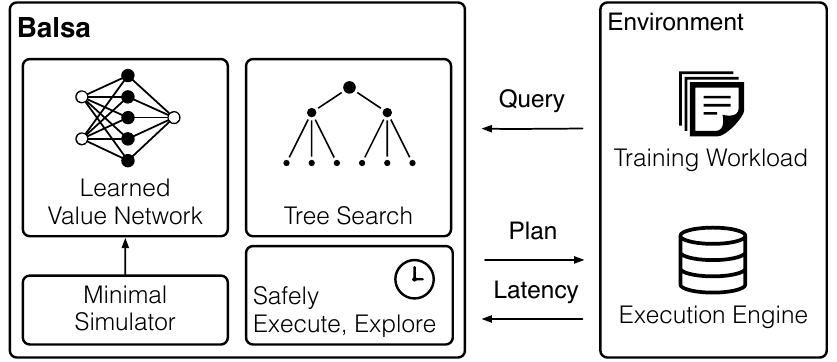}
\vspace{-.2in}
\caption{\sys's architecture.
  \sys learns to optimize queries  by executing plans and observing their latency feedback from an engine.
  \label{fig:arch}}
\vspace{-.3in}
\end{figure}

\sys's goal is to learn to optimize queries for a given dataset and an execution engine.
We assume a training workload is available.
At test time, \sys is asked to optimize unseen queries issued for the same dataset, which can contain new filters and join graphs
that are different from those in the training queries.

\sys
learns by trial and error.
It optimizes the training queries, producing different plans, then executes them on the engine to observe their runtimes.
Based on the runtime feedback, \sys updates itself to correct mistakes and reward good decisions.
As the feedback loop repeats, \sys
gets better at generating good plans.

After training, \sys can be deployed to optimize
an unseen test set of queries.
The agent is evaluated by the performance of training plans produced,
the performance of testing plans produced (i.e., its generalization ability), and its learning efficiency.

Throughout learning, \sys accesses the underlying execution engine only to execute plans and observe their runtimes, and does not learn from an existing  optimizer.
This requirement is informed by the fact that many data systems have execution engines built long before an optimizer becomes available (\secref{sec:intro}).

\vspace{6pt} \noindent {\bf Assumptions.}
We assume the database content is kept static.
Updates to the schema, appends, or in-place updates can be handled by retraining.
This assumption implies that the agent need not solve a learning problem with a shifting distribution.
Another assumption is that
\sys currently optimizes select-project-join (SPJ) blocks.
This is in line with the classical treatment~\cite{systemr} of decomposing a query into simple SPJ blocks and optimizing them block-by-block.

\subsection{Approach}

\sys's architecture is shown in \autoref{fig:arch}.
It
consists of three basic components:
bootstrapping a value network in a minimal cost model,
fine-tuning the value network in real execution,
and using a tree search algorithm  to build query plans.

\vspace{6pt} \noindent {\bf Classical design: cost models $+$ enumeration.}
The classical optimizer design~\cite{systemr} uses an expert-implemented
\emph{cost model} that takes in a plan\footnote{We use ``plans'' to refer to both complete plans and partial subplans.} and outputs a cost estimate: %
\[
C: \textsf{plan} \rightarrow \textsf{cost}
\]
Costs are designed to reflect real execution performance: lower costs should correlate with faster execution.
The optimizer produces plans by \emph{enumerating} candidate plans and scoring them using the cost model. %
For queries with a small number of tables, dynamic programming (DP) is typically used as the enumeration module.

\vspace{6pt} \noindent {\bf RL: value functions $+$ planning.}
Instead of a cost model,
which estimates the immediate cost of a plan, \sys learns a \emph{value function}
that estimates
the \emph{overall} cost/latency of executing a query when the plan is used as a partial step (subplan):
\[
V : (\textsf{query}, \textsf{plan}) \rightarrow \textsf{overall cost or latency}\\
\]
Given a value function, we can use it to optimize queries by building a plan bottom-up.
Consider a query $Q$ joining tables $\{A, B, C, D\}$.
To figure out the best first join to perform, we compare the
overall cost/latency,
i.e., the value, of all valid first joins:
\[
\{A,B,C,D\} \Rightarrow [ V(Q, A\bowtie B)); \, V(Q, A\bowtie C); \, \dots ]
\]
In other words, we use $V$ to score the 2-table joins, which are all partial subplans to complete query $Q$.
The best first join is the one with the lowest $V$ value.
Suppose $A\bowtie C$ is the best among them, then
we can continue the process, scoring all possible second joins:
\[
  \{A\bowtie C,B,D\} \Rightarrow [ V(Q, B \bowtie D); \, V(Q, B \bowtie (A \bowtie C)); \, \dots ]
\]
Continuing such \emph{planning}
leads to a complete query plan.

In contrast to the classical cost model, a value function directly optimizes for the final, overall cost/latency of completing a query---the real objective we care about.
Moreover, a \emph{learned} value function can leverage data to tailor to a target  database and hardware environment, potentially surpassing heuristics.
If the optimal value function $V^\ast$ is known, then planning would produce optimal plans for queries.
Our goal is to approximate $V^\ast$ as accurately as possible. %

\vspace{6pt} \noindent {\bf Learned value networks.}
\sys approximates the optimal value function by training a neural network, $V_\theta (\textsf{query}, \textsf{plan})$ (with parameters $\theta$),
on agent-collected data. %
The two inputs to the network are featurized into \emph{query features} (encoding joined tables and filters) and \emph{plan features} (encoding the tree structure of the plan and each node's operator type), respectively. %

We learn the value function in two stages.
First, we learn parameters $\theta_{\text{sim}}$ in a fast simulation environment backed by a minimal cost model.
Next, we initialize parameters $\theta_{\text{real}} \leftarrow \theta_{\text{sim}}$ and start fine-tuning the value function in real execution.
The two stages produce the value networks\footnote{For notational convenience, throughout the paper we use  \vsim and \vreal  to refer to the simulation and real-execution models $V_{\theta_{\text{sim}}}$ and $V_{\theta_{\text{real}}}$, respectively.}:
\begin{align*}
V_\text{sim} &: (\textsf{query},\textsf{plan}) \rightarrow \textsf{overall cost}\\
V_\text{real}&: (\textsf{query},\textsf{plan}) \rightarrow \textsf{overall latency}
\end{align*}
After training,  \vreal is used with planning to optimize new queries.

\vspace{6pt} \noindent {\bf Step 1: bootstrapping from a minimal cost model (\secref{sec:sim_learning}).}
\sys starts learning in a ``simulator'' of query optimization, i.e., a cost model.
The key advantage of using a simulator is that \emph{the agent can learn about disastrous plans without executing them} in the initial phase of learning.
The agent bootstraps initial knowledge against an {inaccurate  but fast-to-query} cost model,
which provides rapid feedback (cost estimates) for the agent. %
The cost model is generic and does not model the target engine or hardware.

To train the simulation model \vsim, we use a data collection procedure (e.g., DP) to enumerate plans for the training query set and ask the simulator for costs.
Each query can yield thousands of training data points, eventually producing a sufficiently large dataset, $\mathcal{D}_\text{sim} = \{(\textsf{query}, \textsf{plan}, \textsf{overall cost})\}$.
\vsim is then trained on this dataset in a standard supervised learning fashion.%

\vspace{6pt} \noindent {\bf Step 2: fine-tuning in real execution (\secref{sec:real_learning}).}
Next, we transfer the value function from doing well in the simulator
to excelling in the real execution environment.
The second stage starts by initializing the real-execution model from the trained simulation model: \vreal $\leftarrow$ \vsim.
The fine-tuning of \vreal is performed in iterations of query executions and model updates.
In each iteration, \sys uses its current \vreal to optimize training queries; these plans are executed with their latencies measured. %
\sys then updates its \vreal on these collected data to make its latency predictions more accurate.

A key challenge of learning in real execution is mitigating slow plans. %
We address this as follows.
By initializing from \vsim,  \sys's behavior in iteration 0 would be much better than random initialization (which amounts to picking plans randomly).
After iteration 0, \sys uses timeouts (determined by earlier runtimes) to early-terminate slow plans (\secref{sec:timeout}) and
also employs safe exploration (\secref{sec:exploration}).

\vspace{6pt} \noindent {\bf Planning  with tree search.}
\sys uses
tree search planning
on top of the learned value function to optimize queries.
The learned \vreal guides the search towards the promising regions of the plan space.
As \vreal becomes more accurate, better plans can be found.

There are many tree search algorithms  with different complexity-optimality tradeoffs: from greedy planning, to
advanced planning algorithms such as Monte Carlo tree search.
We opt for a middle ground
by using a simple beam search (\secref{sec:beam-search}).

\vspace{6pt} In the next sections, we describe \sys's components in detail.
\section{Bootstrapping From Simulation}
\label{sec:sim_learning}
The first stage of training aims to rapidly impart basic knowledge to the agent, before it starts learning in long-running real executions. %
We achieve this by bootstrapping \sys in a minimal simulator, i.e., a cost model.
It ``simulates'' query optimization in that query plans are not actually executed.
Instead, the agent issues a large amount of plans to the simulator, which can quickly return cost estimates (rather than measuring their runtimes) as feedback.
\noindentparagraph{\bf Why is a simulator necessary?}
The search space for a query is vast and disastrous execution plans are abundant~\cite{leis2015good}.
Unfortunately, disastrous plans can stall learning progress: an agent may wait for a long time
for a slow plan to complete execution, before learning that it is a bad action (if it ever finishes).
This property is in direct contrast to other RL use cases such as games.
In game environments (e.g., Go, chess, Atari), bad moves typically cause a game to end \emph{sooner}, as the opponent can exploit the agent's mistakes.

A randomly initialized RL agent without training in simulation can quite easily stumble upon such disastrous plans, especially in the early stage of learning.
We show this with a simple experiment: %
we randomly initialize 6 agents without simulation learning, and task them with optimizing 94 queries from the Join Order Benchmark (detailed setup described in \secref{sec:eval:setup}). %
Plans produced by
the median random agent execute $45\times$ slower in workload runtime than those produced by an expert optimizer, \pg.  The slowest agent is $79\times$ slower than the expert (2.5 hours vs. 2 minutes).

Next, we describe the specific choice of cost model employed.

\subsection{A Minimal Simulator}
\label{sec:sim_cost_model}
\sys uses a minimal, {logical plan-only} cost model, which captures the general principle that ``fewer tuples lead to better plans''.
{It is \emph{minimal}, because it is free of any prior knowledge about the execution engine and physical operators (e.g., merge vs. hash join).}

Formally, we use the \cout cost model~\cite{cluet1995complexity}:%
\begin{equation*}
\label{eq:mincardcost}
C_{out}(T) =
\begin{cases}
|T| &\text{if $T$ is a table/selection} \\
|T| + C_{out}(T_1) + C_{out}(T_2) &\text{if $T = T_1 \bowtie T_2$}
\end{cases}
\end{equation*}
where $|T|$ denotes the estimated cardinality of a table (with filters taken into account) or a join, obtained from a
cardinality estimator {(\secref{sec:sim-discussion})}.
This cost model estimates the cost of a query plan simply by summing up the estimated result sizes of all operators.

\noindentparagraph{\bf Tradeoffs of a minimal simulator.}
We choose a minimal cost model to bake in as little prior knowledge as possible.
The goal of simulation learning is to steer the agent away from definitively disastrous plans (when it starts the real execution phase), not to instill expert knowledge.
It is also \emph{generic}: {by not modeling physical details, it can be used to bootstrap \sys optimizing for any engine}.

Due to its simplicity, the cost model is inherently inaccurate.
\sys will learn to fill in missing knowledge and correct inaccuracy when fine-tuning in the real execution phase (\secref{sec:real_learning}).
As we will show in \secref{sec:ablation-sim},
while \sys can leverage pre-engineered, more sophisticated cost models to accelerate training, they are not required for \sys to reach expert-level performance.

\subsection{
Simulation Data Collection
}
\label{sec:data-collection}
Given a simulator, we extract as much knowledge from it as possible by applying a
batched
data collection procedure. %
The output is the simulation dataset, $\mathcal{D}_\text{sim} = \{(\textsf{query}, \textsf{plan}, \textsf{overall cost})\}$, which is used to train the value network \vsim.
Specifically, we use dynamic programming (also used by DQ~\cite{dq}) to collect data. %

\begin{figure}[tp]
\centering
\includegraphics[width=\columnwidth]{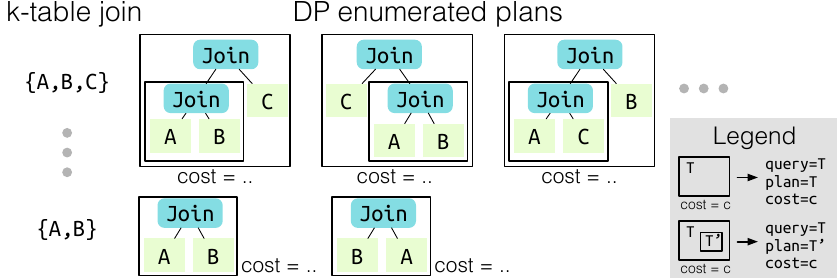}
\vspace{-.25in}
\caption{%
  Simulation data collection and augmentation.
  For each $k$-table join in DP, \sys collects and augments all its enumerated plans.
  Each bordered box yields a collected data point (see legend).
  \label{fig:data-collection}}
\vspace{-.05in}
\end{figure}

\noindentparagraph{\bf Enumerating plans using dynamic programming.}
For each query in \sys's training workload, we run the classical Selinger~\cite{systemr} bottom-up DP with a bushy plan space.
It starts by enumerating the best plans for all valid 2-table joins, composed out of base table scans, then enumerating 3-table joins, etc.
Each enumerated plan $T$ will get a cost estimate $C$ from the cost model\footnote{\sys enumerates physical plans for \cout, which will ignore the differences between physical joins/scans and treat them as logical operators.}, generating a data point
\textsf{(query=$T$, plan=$T$, overall cost=$C$)},
where \textsf{query=$T$} denotes the original query restricted to the tables/filters of $T$.
This data point undergoes a \emph{data augmentation} procedure, described below, to  yield a list of training data points to be added into $\mathcal{D}_\text{sim}$.

The data collection is \emph{high-throughput}: data is generated from  \emph{all} enumerated plans, %
not just from the
set of optimal plans in the final DP results.
This means that some suboptimal plans (under the cost model) are included, which increase data variety and aid learning.
\autoref{fig:data-collection} illustrates the data collection procedure.

However, DP's runtime may become too large for queries joining many tables.
Hence we skip collecting data from queries with $\geq n$ tables (we set $n = 12$).
Alternative strategies can also be applied. For example, DQ proposes a partial DP scheme where the first $j$ levels of DP are run and the rest of the levels are planned greedily.

\noindentparagraph{\bf Data augmentation.}
\sys employs a data augmentation technique proposed by DQ,
where multiple data points are generated from a single enumerated plan.
Specifically, given a \textsf{(query=$T$, plan=$T$, overall cost=$C$)}, each subplan $T'$ of $T$ will yield a distinct data point with the same ``overall query'' $T$ and the same cost: $\{\textsf{(query=$T$, plan=$T'$, overall cost=$C$)}: \forall T' \subseteq T \}$.
This technique significantly enriches the dataset $\mathcal{D}_\text{sim}$ in quantity and variety.

\textbf{Interpretation.} In RL terms, the augmentation reflects that all states (the subplans) in a trajectory (the overall query/final plan) share the same return, because intermediate rewards are defined to be 0 and terminal rewards are the negative costs of final plans.

\subsection{Discussion}
\label{sec:sim-discussion}

We found simulation learning to be  highly effective.
At the start of \secref{sec:sim_learning}, we performed a simple experiment illustrating an up to $79\times$ gap between randomly initialized (i.e., no bootstrapping) agents and an expert optimizer.
Now, with simulation bootstrapping, agents significantly shorten this gap  to only $5.8\times$ slower than the expert at max---all without performing any real execution. 

{
\noindentparagraph{\bf {Cardinality estimator}.}
The simulator needs a cardinality estimator.
As mentioned in \secref{sec:intro}, we pick \pg's estimator for its simplicity
(per-column histograms;
heuristically assumes independence for joins;
``magic constants'' for complex filters)~\cite{leis2015good}.
\sys does \emph{not} learn from \pg's optimizer (costs or plans).
}

{%
We use an {existing}, textbook-style estimator for convenience,
\emph{not to rely on it for good performance}.
In fact, most of \sys's quality improvements are learned after the simulation stage (\secref{sec:eval:perf}, \secref{sec:discussions}).}

\noindentparagraph{\bf Alternative cost models.}
While \sys advocates for a minimal simulator,
more prior knowledge can be plugged in by the user, if desired.
Other cost models may include progressively more physical operator knowledge (e.g., the $C_{mm}$ cost model~\cite{leis2015good} for in-memory settings).
New query engines optimizing for different objectives (e.g., lower memory footprint) may either bootstrap \sys with \cout (its fewer-tuples-are-better principle generally applies), or develop another minimal cost model tailored to the objective.

\section{Learning from Real Execution}
\label{sec:real_learning}
Simulation learning imparts basic knowledge to the agent.
But no
{simulators} can perfectly reflect the nuances of the real execution environment.
Therefore, we fine-tune the agent through query executions in the real environment.

\subsection{Reinforcement Learning of the Value Function}
\label{sec:rl-algo}
\sys learns the real-execution value network, $V_\text{real}(\textsf{query},\textsf{plan}) \rightarrow \textsf{overall latency}$, using reinforcement learning.
The basic idea is that the agent iteratively uses its current value network to optimize queries and runs them, then uses the latency feedback to improve itself.
As this feedback loop runs, more execution data is collected,
and the agent's \vreal becomes  better at generating good plans.

Concretely, we start with \vreal initialized%
\footnote{Predictions naturally change from the scales of costs to latencies through fine-tuning.}
from \vsim and an empty real-execution dataset, $\mathcal{D}_\text{real} =\emptyset$.
Each iteration of learning consists of an execute and an update phase.

\textbf{\emph{Execute.}}
The agent uses the current \vreal to optimize each training query $q$, producing an execution plan $p$.
(Planning will be described in \secref{sec:beam-search}.)
Each plan is executed on the target engine with its latency $l$ measured.
This results in one data point, \textsf{(query=$q$, plan=$p$, overall latency=$l$)}, which then undergoes the same subplan data augmentation discussed in~\secref{sec:data-collection} to yield a list of data points:
\[
  \mathcal{D}_\text{real} \mathrel{{+}{=}} \{(\textsf{query}=q, \textsf{plan}=p', \textsf{overall latency}=l) : \forall p' \subseteq p\}
\]

\textbf{\emph{Update.}}
\sys uses the collected data to improve its \vreal.
We perform stochastic gradient descent (SGD) with an L2 loss between predicted and true latencies.
Thus, mispredictions are corrected and good predictions are reinforced.
Data points $(q,p,l)$ are sampled from  $\mathcal{D}_\text{real}$.
However, model outputs $V_\text{real}(q,p)$ are updated not towards $l$, but towards the \emph{best latency obtained so far} of  query $q$ that involves subplan $p$---%
{a previously proposed technique~\cite{neo}}.
The latency label correction is motivated as follows.
Consider query $q$ joining tables $A, B, C, D$.
Subplan $p = $~\textsf{Join}$(A,B)$
may have appeared in
two executions, one with $C$ joined next and one with $D$ joined next.
They may have wildly different latencies, say 1 vs. 100 seconds. %
As we wish to minimize latency, we define the lower latency $l=1$ as the value of subplan $p$, because $p$ could have made $q$ run this fast.
The best latencies so far are calculated from the entire $\mathcal{D}_\text{real}$.

\vspace{6pt}
Thus, data collection and value function improvement alternate.
The algorithm can be thought of as either value iteration~\cite{rlbook} or expert iteration~\cite{anthony2017thinking}, and variants of it have  been recently applied in prior work in query optimization~\cite{neo} (which, different from \sys's updates, resets and retrains the value network across iterations)
, theorem proving~\cite{gptf}, and compute schedule optimization~\cite{adams2019learning}.

\noindentparagraph{\bf On-policy learning.}  %
\sys employs a novel optimization on top of the algorithm above by using \emph{on-policy learning}.
Updates to \vreal are performed only on the data points generated by the current \vreal. %
In other words, SGD is performed on data points $(q, p, \_)$ sampled from the most recent iteration of the dataset, $\mathcal{D}_\text{real}$, but not from its entirety.
The latter would yield data from many iterations ago and is hence off-policy. %
Label correction still utilizes the entire dataset.

Intuitively, the most recent data points generally are the most surprising to the agent and have faster latency labels, so it should be beneficial to focus on them.
Indeed, we find on-policy learning to significantly accelerate learning, by reducing the number of SGD steps per iteration, and improve the plan variety and performance of \sys (\secref{sec:eval:ablation:on-pol}).
On-policy learning makes \sys's training more than {$9.6\times$} faster when compared
to
Neo~\cite{neo}, {a prior state-of-the-art method}, which employs a full retraining scheme instead
(\secref{sec:compare-to-neo}).
We hypothesize that {this technique} may  also improve other applications of value functions that predict runtimes.

\subsection{Plan Search}
\label{sec:beam-search}

With the learned value network, \sys uses a simple (best-first) beam search to produce execution plans for a given query.

Beam search operates on \emph{search states}, each a set of partial plans for the query.
The search starts with a root state that contains all tables (scans) in the query.
A beam of size $b$ stores search states to be expanded, sorted by their predicted latencies\footnote{\vreal takes a $(\textsf{query}, \textsf{plan})$ as input, while a search state is a set of partial plans for the same query.  To score the latter, we define $V(\textsf{state}) \equiv \max_{\textsf{plan} \in \textsf{state}} V(\textsf{query}, \textsf{plan})$. Intuitively, it reflects that a state's latency is at least the maximum overall latency a subplan is predicted to take.}. %
At each step, the best search state is popped from the beam, and all available actions are applied to produce children states.
Each action joins two eligible plans in the current state with a physical join operator assigned, as well as assigning scan operators if either side is a table.
As a search state is a set of partial plans (joined relations and non-joined tables), applying actions to it will lead to at least one complete plan.

Then, all resulting children states are scored by the value network \vreal and added to the beam, which keeps the top $b$ states only.
In this way, the learned value network \emph{guides the search} to focus on the more promising regions of the plan space.
Beam search terminates when $k$ complete plans are found.
\sys uses $b=20$ and $k=10$.

\noindentparagraph{\bf Top-$k$ plans and exploration.}
Beam search is not guaranteed to return globally optimal plans, and better plans may be found later in the search.
We thus continue searching until $k$ complete plans are found.
At test time, the best plan out of this list is emitted.

Interestingly, at training time, obtaining a list of plans enables a simple exploration technique on top.
We treat all of these plans as having reasonable optimality---so that it should be safe to explore among them---and prioritize choosing the unseen plans as beam search outputs.
This technique is discussed in \secref{sec:exploration}.

\subsection{Safe Execution via Timeouts}
\label{sec:timeout}

A unique challenge in query optimization is the proliferation of expensive plans in a vast search space, even when fast plans exist.
When \sys learns by trial and error from real executions, it can encounter long-running plans with unacceptably high latencies.

\sys addresses this challenge by  applying \emph{timeouts}, a classical idea in distributed systems.
Since training proceeds in iterations, earlier execution runtimes of the same training workload are known and can be used to bound future iterations.

Key to this mechanism is how to pick the initial timeout.
Fortunately, simulation learning allows us to assume that when the real execution starts, the first ever plans produced for a set of training queries have reasonable (albeit suboptimal) latencies.

\noindentparagraph{\bf Timeout policy.}
During iteration 0's execute phase (just after simulation learning), the plans are allowed to finish execution in their entirety---simulation learning is assumed to yield a non-disastrous starting point.
Let the maximum per-query runtime recorded be $T$.

For iteration $i > 0$,
a timeout of $S \times T$ is applied for all agent-produced plans, where $S$ is a ``slack factor''.
By definition of $T$, for any training query there exists a plan that can finish execution in time $T$.
The slack's purpose is to give some extra room  and account for runtime variance (\sys uses $S=2$).

If a plan has been executing longer than the current timeout, it is terminated early, since it would be slower than earlier found plans for the same query anyway.
It gets assigned a large label\footnote{We use 4096 seconds throughout. It can also be set as some multiple of iteration 0's maximum per-query runtime.} instead of its true, unknown latency. Such large labels serve to \emph{discourage} and steer the agent away from similar plans in future iterations.

Timeouts are progressively tightened.
If an iteration finishes with a maximum per-query runtime $T' < T$, then the next iteration's timeout is tightened to $S \times T'$.
This progression ensures that the timeout is neither too small, which prevents progress, nor too large, which wastes efforts.
It generates an \emph{implicit learning curriculum} for the agent with just-about-right difficulties.

In sum, we found the timeout mechanism to significantly accelerate learning.
It bounds the runtime of each iteration's execute phase and eliminates unexpected stalls, thereby achieving \emph{safe execution}.

\section{Safe Exploration in Real Execution}
\label{sec:exploration}

While an RL agent exploits its past experience for good performance, it must also explore new experience to escape local minima.
To achieve this, an exploration strategy can be used. %

However, the abundance of slow plans, a unique characteristic of query optimization,
additionally requires \emph{safe exploration}, i.e., disastrous plans be avoided.
Random plans sampled from the search space are slow~\cite{leis2015good}, and choosing to explore them would again stall learning.
In our early experiments, a basic $\epsilon$-greedy strategy (for each  training query, with a small $\epsilon$ probability a random plan is sampled, a la QuickPick~\cite{waas2000join})
often selected inferior plans that led to timeouts, slowing down the discovery of better plans and learning.

To achieve safe exploration, \sys proposes a simple \emph{count-based exploration} technique.
In essence, this family of methods encourages an agent to explore a less-visited state or execute a less-chosen action.
We instantiate this principle in the following way.

\begin{figure}[tp]
\centering
\includegraphics[width=\columnwidth]{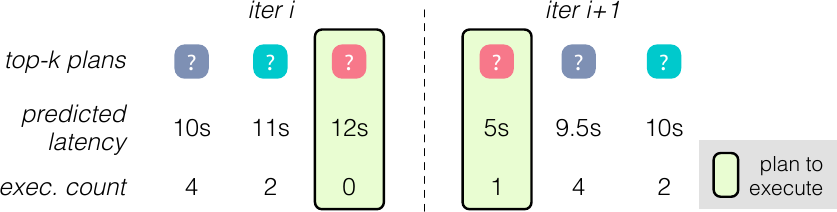}
\vspace{-.25in}
\caption{%
  Safe exploration.
  For a training query, \sys prioritizes running the unseen plans of the top-$k$ plans from tree search (exploration).
  If all seen, the predicted-best plan is chosen (exploitation).
  \label{fig:safe-explore}}
\vspace{-.25in}
\end{figure}

\noindentparagraph{\bf Count-based exploration for beam search.}
Our goal is to provide a ``trust region'' of reasonable plans for the agent to explore.
To do so, beam search is asked to return top-$k$ plans, sorted by ascending predicted latencies, rather than the single best plan found.
Instead of executing the best plan (i.e., with the lowest predicted latency),
we execute \emph{the best unseen plan} of this list.
If all top-$k$ plans have been previously executed---indicating sufficient exploration---\sys resorts to exploitation by executing the predicted-cheapest plan.
The visit counts of plans are cached by a hash table, which adds low overheads, as past executions are already stored in $\mathcal{D}_\text{real}$.
\autoref{fig:safe-explore} illustrates this technique using example statistics ($k=3$).

Intuitively,  all of the top-$k$ plans are
\emph{probably good}
(since they are produced by  value network-guided beam search), so they should not be chosen strictly by their predicted latencies (which are imperfect estimates).
Therefore, executing novel, unseen plans in this ``trust region'' is both safe and exploratory. %

\section{Diversified Experiences}
\label{sec:generalization}
{
  For learned query optimizers, robustly optimizing unseen queries is essential.
  To further enhance \sys's generalization performance, we introduce a simple method, \emph{diversified experiences}.

\noindentparagraph{\bf Problem: mode diversity.}
As a value network is used to guide plan search, an agent tends to only experience plans preferred by its value network, and may gradually
converge to plans
with similar characteristics, or a ``mode''~\cite{wf_mode}.
For example, if hash and loop joins are equally effective for a workload, an agent may learn to heavily use hash joins, while another may prefer loop joins.
Either agent can output good plans, as both operators are effective, but they may lack the knowledge about plans that prefer alternative operators or shapes.
(While exploration increases plan variety, the new plans are still relatively confined to a single agent's mode.)
Low mode diversity can hinder an agent's generalization to highly distinct, unseen queries that require unfamiliar modes to be optimized well.

\begin{figure}[tp]
\centering
\includegraphics[width=\columnwidth]{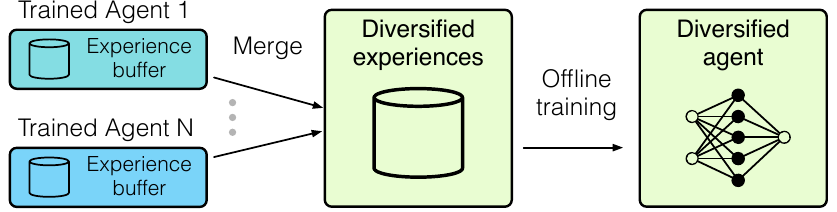}
\vspace{-.25in}
\caption{%
  Diversified experiences. A more robust agent is produced by retraining on the experiences collected from different agents.
  \label{fig:diversified}}
\end{figure}

\noindentparagraph{\bf Diversified experiences.}
To enhance generalization, we propose simply merging the experiences ($\mathcal{D}_\text{real}$) collected by several independently trained agents (with different random seeds),
and retraining a new
agent on top
without any real execution.
\autoref{fig:diversified} illustrates this process.
Our insight is that this \emph{diversified experience covers multiple modes}. Thus, training on it produces a more robust value network that generalizes better.

\vspace{-.05in}
\begin{table}[htp]\centering\small%
  \caption{{
    Diversifying experiences: number of data collection agents vs. number of unique plans after merging.
    Agents have highly diverse experiences.
    Trained on 113 JOB queries (details in \secref{sec:eval:setup}).
  }
    \label{table:reborn-data-diversity}}
\vspace{-.1in}
\begin{tabular}{@{} l r r r @{}}
\toprule
  {Num. Agents} & 1 & 4 & 8 \\
  \midrule
{Num. Unique Plans} & 27K (1$\times$) & 102K (3.8$\times$) & 197K (7.3$\times$)\\
\bottomrule
\end{tabular}
\vspace{-.05in}
\end{table}

\autoref{table:reborn-data-diversity} confirms this insight: the number of unique plans grows almost linearly as the number of agents, showing that the plans experienced by different agents
are indeed highly diverse.
We find this simple method effective (\secref{sec:eval-generalization}), offering a way to trade more compute, when available, for better performance.}

\section{\sys Implementation}
\label{sec:impl}

In this section we describe \sys's detailed training setup.
At a high level, to operate \sys on a new engine it  needs the following:
\begin{itemize}[leftmargin=*]
\item An execution environment (executes plans; support for timeouts).
\item Definition of the search space (the set of query operators and the rules to compose them).
\end{itemize}

\begin{figure}[t]
\centering
\includegraphics[width=\columnwidth]{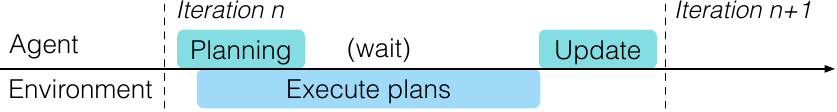}
\vspace{-.25in}
\caption{Pipelining agent planning and remote query execution.
  \label{fig:training_timeline}}
\vspace{-.1in}
\end{figure}

\noindentparagraph{\bf {Optimizations}.}
{We optimize training by parallel data collection, plan caching, and pipelining.}
Query executions are dispatched to a pool of identical virtual machines each running an instance of the target database, using Ray~\cite{ray}. Each VM runs one query at a time to prevent interference.
A plan cache is used
so that reissued plans have their prior runtimes quickly looked up and can skip re-execution.
Planning and remote query execution in each iteration are pipelined (\autoref{fig:training_timeline}): as soon as tree search (run by the main agent thread) finishes planning a training query, the output plan is sent for remote execution, and then planning for the next query starts.
The two stages thus overlap.
The agent waits for all plans to finish before performing value network updates.

\noindentparagraph{\bf Value network details.}
The value networks, \vsim and \vreal, {are implemented as simple tree convolution networks~\cite{neo}} (0.7M parameters, or 2.9MB).
We also experimented with implementing them using a Transformer~\cite{transformer} early on;
this was found to be similarly effective but had higher computational costs.
When training or updating the value networks, we sample 10\% of  experience data
as a validation set for early stopping.
The inputs to the value network, \textsf{query} and \textsf{plan}, are encoded as follows.
{Each \textsf{plan} has the same encoding as Neo~\cite{neo}.}
A \textsf{query} is featurized as a vector $\textsf{[} \textsf{table} \rightarrow \textsf{selectivity}\textsf{]}$ where each slot corresponds to a table and holds its estimated selectivity {(\secref{sec:sim-discussion})}.
Absent tables' slots are filled with zeros.
This encoding is simpler than {both Neo and DQ~\cite{dq}}.

\section{Evaluation}
\label{sec:eval}
We conduct an in-depth evaluation of \sys. Our key findings are:
\begin{itemize}[leftmargin=*]
\item {Learning by trial and error, \sys generates better execution plans that run up to 2.1--2.8$\times$ faster in workload runtime than two expert optimizers, \pg and ``\dbms''\footnote{A leading commercial DBMS. We anonymize its name due to its licensing terms~\cite{read2006dewitt}.} (\secref{sec:eval:perf}).}

\item {\sys takes a few hours to surpass the experts and a few more hours to reach peak performance on the tested workloads (\secref{sec:eval:perf}).} %
\item {\sys outperforms learning from expert demonstrations~\cite{neo}, a prior state-of-the-art approach, despite not learning from an expert optimizer (\secref{sec:compare-to-neo}).
  We also identify \emph{poor generalization} as a potential failure mode in this prior method.} %
\item {Diversified experiences significantly enhance generalization, including to queries with highly distinct join templates (\secref{sec:eval-generalization}).}
\item {\sys learns \emph{novel} preferences of operators and plan shapes (\secref{sec:behavior}).}
\end{itemize}
Additionally, we conduct detailed ablation studies to understand the effect of \sys's design choices in \secref{sec:ablation}.

\subsection{Experimental Setup}
\label{sec:eval:setup}
We use the following workloads, in each of which \sys is trained on a set of training queries and tested on a set of unseen queries:

\noindentparagraph{\bf Join Order Benchmark (JOB)} contains 113 analytical queries designed by Leis \etal \cite{leis2015good} to stress test query optimizers over a real-world dataset from the Internet Movie Database.
The queries involve complex joins and predicates, ranging from 3-16 joins, averaging 8 joins per query.
We benchmark against two train-test splits, each with 94 training and 19 test queries:
\begin{itemize}[leftmargin=*]
\item Random Split (denoted as ``JOB''):  a randomly sampled split. %
\item Slow Split (denoted as ``JOB Slow''): the test set consists of the 19 slowest-running queries when planned by {an} expert optimizer. %
\end{itemize}
Random Split tests an average situation, while Slow Split evaluates when the test queries run maximally slower than the train queries.

\noindentparagraph{\bf TPC-H} is a standard analytical benchmark where data and queries are generated from uniform distributions.
We use a scale factor of 10. %
We use 70 queries for training and 10 queries as the test set\footnote{For TPC-H, we use templates 3, 5, 7, 8, 12, 13, 14 for training and template 10 for testing, with 10 queries generated per template.
We avoid the templates with advanced SQL features (views, sub-queries) due to a limitation in the \textsf{pg\_hint\_plan} extension.}.

\noindentparagraph{\bf {Expert baselines and engines.}}
{We compare with the optimizers of two mature expert systems: \pg (12.5; open-source) and \dbms (a leading commercial DBMS; anonymized~\cite{read2006dewitt}).
  For each expert, we compare \sys's plans with its optimizer's plans executed on that same engine. %
\sys's plans are injected by hints~\cite{pghint}.}

We use Microsoft Azure VMs with 8 cores, 64GB RAM, and SSDs.
Training is done on a NVIDIA Tesla M60 GPU. %
We configure \pg with 32GB shared buffers and cache size, 4GB work
memory, and GEQO disabled---settings similar to Leis \etal \cite{leis2015good}.
{We optimize \dbms extensively by following its tuning guides.}

\sys is trained for 500 iterations on the JOB workloads and 100 iterations on TPC-H due to its smaller search space. \sys uses all components and default values discussed in prior sections.  %

{\bf {Expert performance\footnote{\pg runtimes (train/test): JOB 115s/24s; JOB Slow 44s/98s, TPC-H 452s/49s. {We do not disable nested loop joins as suggested by Leis \etal, because with indexes created, this change actually made the expert run JOB 60\% slower.}}.}} We follow the guidance in Leis \etal \cite{leis2015good} to create all primary and foreign key indexes to make our baselines run JOB much faster than that of prior work~\cite{neo,skinnerdb_sigmod}.
This also makes the search space more complex and challenging.

\noindentparagraph{\bf Metrics.}
We repeat each experiment 8 times and report the median metric, unless specified otherwise.
{In train/test curves, we show the \emph{entire min/max ranges} in shaded areas.}
Workload runtime is defined as the sum of per-query latencies.
When reporting normalized runtimes, they are calculated with respect to the expert's runtimes.

\subsection{\sys Performance}
\label{sec:eval:perf}
We begin with end-to-end results, answering the following:
\begin{itemize}[leftmargin=*]
\item What is the performance of \sys on training and test queries?
\item How many hours (and executions) does \sys need to surpass expert performance and reach its peak performance, respectively?
\end{itemize}
\begin{figure}[tp]
\centering
\includegraphics[width=\columnwidth]{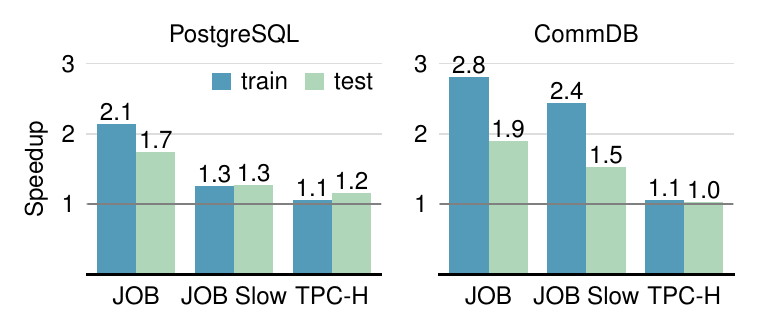}
\vspace{-.35in}
\caption{{\sys's performance on \pg (left) and \dbms (right):
    workload speedups achieved by \sys plans over plans from the respective expert optimizer.
    Each bar is the median of 8 runs.
  }
  \label{fig:perf}}
\vspace{-.1in}
\end{figure}
\noindentparagraph{\bf Performance.}
\autoref{fig:perf} summarizes \sys's overall performance.
{On all workloads, \sys is able to start from a minimal cost model and learn to surpass the expert optimizers by a sizable margin.}

On \pg, \sys achieves a $2.1\times$ training-set speedup on JOB, $1.3\times$ on JOB Slow, and $1.1\times$ on TPC-H.
While speedups on test sets slightly trail behind the training set speedups, \sys can still produce faster execution plans than the expert (e.g., $1.7\times$ faster on JOB).
This shows that \sys can generalize to unseen queries. %

{\sys also outperforms \dbms's optimizer. The speedups are higher---1.1--2.8$\times$ for train and 1.0--1.9$\times$ for test sets---because \dbms
  allows
  a much smaller search space than \pg
  by not exposing
  bushy hints. (We estimate it to be 1000$\times$ smaller for an average-sized JOB query, counting plan shapes and operators.) \sys thus explores the smaller search space more comprehensively.}

\noindentparagraph{\bf {Runtime of simulation learning.}}
{\autoref{table:sim} shows simulation is data-rich and takes dozens of minutes. As it is a small fraction of real execution learning's duration, we focus on the latter next.}

\begin{figure*}[t]
     \centering
     \begin{subfigure}[b]{0.472\textwidth}
         \centering
          \includegraphics[width=.495\columnwidth]{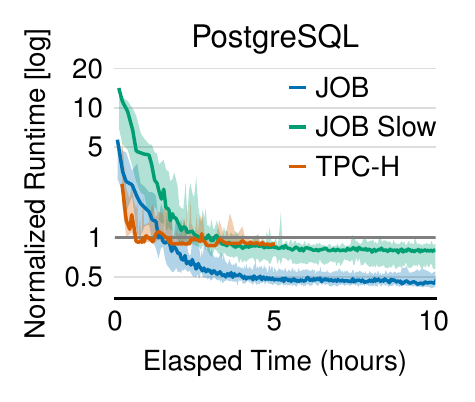}
          \includegraphics[width=.495\columnwidth]{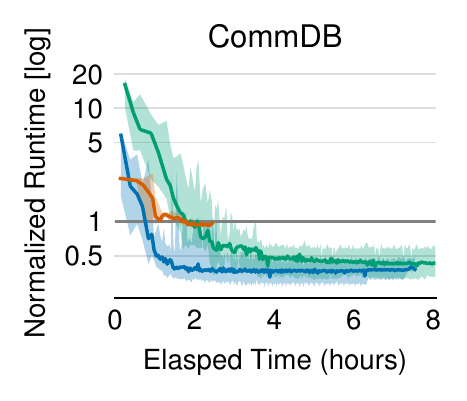}
\vspace{-.25in}
         \caption{Wall-clock efficiency\label{fig:wallclock-curve}}
     \end{subfigure}\hfill
     \begin{subfigure}[b]{0.472\textwidth}
         \centering
          \includegraphics[width=.495\columnwidth]{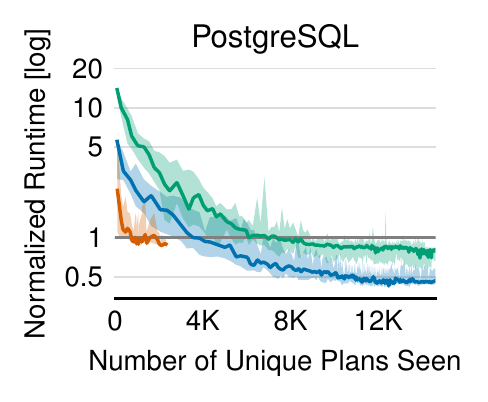}
          \includegraphics[width=.495\columnwidth]{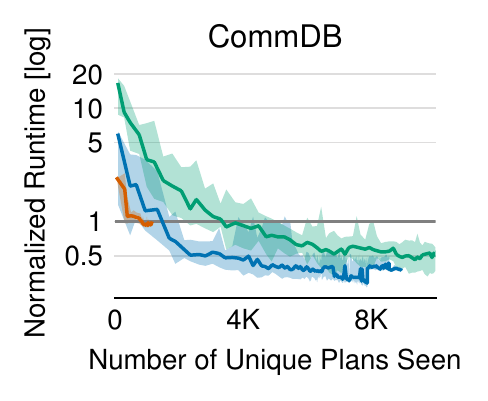}
\vspace{-.25in}
         \caption{Data efficiency\label{fig:sample-curve}}
     \end{subfigure}\hfill
\vspace{-.10in}
\caption{{Learning efficiency of \sys.} Normalized runtime of training queries (log scale) vs. (a) elapsed time and (b) number of executed plans.
\label{fig:learning-efficiency}}
\vspace{-.25in}
\end{figure*}

\begin{figure}[t]
     \centering
     \begin{subfigure}[b]{0.236\textwidth}
         \centering
          \includegraphics[width=\columnwidth]{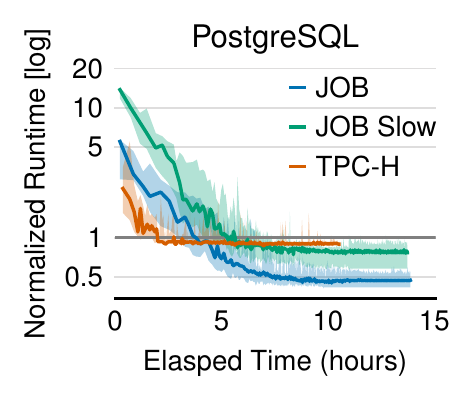}
\vspace{-.23in}
     \end{subfigure}\hfill
     \begin{subfigure}[b]{0.236\textwidth}
         \centering
          \includegraphics[width=\columnwidth]{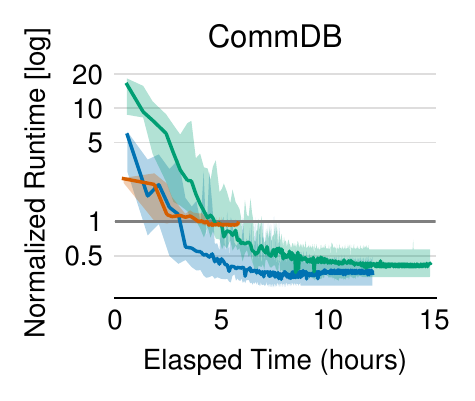}
\vspace{-.23in}
     \end{subfigure}\hfill
\vspace{-.10in}
\caption{{Wall-clock efficiency, non-parallel training mode.}\label{fig:wallclock-non-par}}
\vspace{-.10in}
\end{figure}

\begin{figure}[t]
     \centering
     \begin{subfigure}[b]{0.236\textwidth}
         \centering
          \includegraphics[width=\columnwidth]{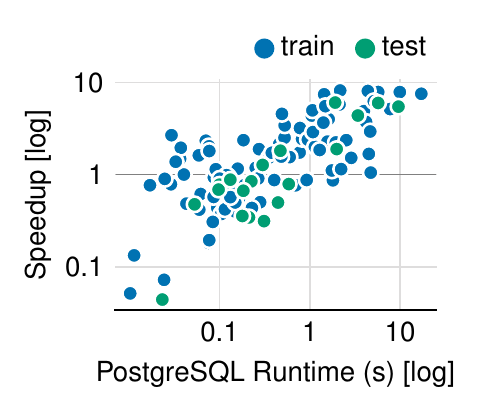}
\vspace{-.23in}
         \caption{JOB\label{fig:speedup-job}}
     \end{subfigure}\hfill
     \begin{subfigure}[b]{0.236\textwidth}
         \centering
          \includegraphics[width=\columnwidth]{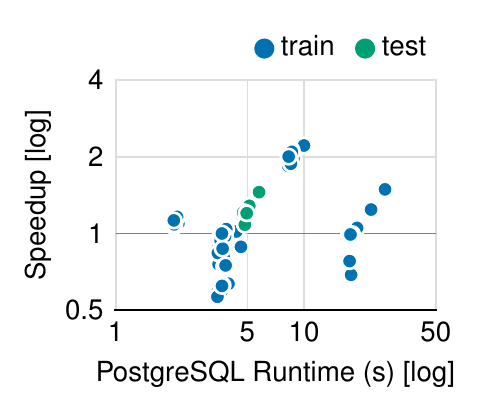}
\vspace{-.23in}
         \caption{TPC-H\label{fig:speedup-tpch}}
     \end{subfigure}\hfill
\vspace{-.10in}
\caption{{Breakdown of \sys's per-query speedups.}
Speedup of each query (log scale) vs. \pg expert runtime (log scale).
\label{fig:speedup}}
\vspace{-.10in}
\end{figure}

\vspace{-.05in}
\begin{table}[htp]\centering\small%
  \caption{Simulation learning efficiency: sizes of simulation datasets, time to collect data (in minutes), and time to train.
    Train times differ due to early stopping.
    Means $\pm$ standard deviations are shown.
    \label{table:sim}}
\vspace{-.1in}
\begin{tabular}{@{} l r r r @{}} \toprule
{Workload} & {Size} & {Collection time (min.)} & {Train time (min.)} \\   \midrule

JOB & 516K & $6.8 \pm 0.1$ & $24 \pm 8$ \\

JOB Slow & 551K & $7.6 \pm 0.1$ & $28 \pm 10$ \\

TPC-H & 12K & $1.1 \pm 0.01$ & $1.0 \pm 0.2$ \\
\bottomrule
\end{tabular}
\vspace{-.05in}
\end{table}

\noindentparagraph{\bf Learning efficiency.}
Figure \ref{fig:learning-efficiency} shows the training performance of \sys as a function of elapsed time and the number of distinct query plans executed.
(The latter is called data/sample efficiency in RL terms, as each execution is an interaction with the environment.)

\emph{\bf Wall-clock efficiency.}
\autoref{fig:wallclock-curve} shows \sys's wall-clock efficiency during the real execution stage.
\sys starts off several times slower than the {experts}---this is the performance after bootstrapping from a simple simulator.
With just a few hours of learning, \sys matches the {experts' performance (on \pg: 1.4 hours for JOB, 2.5 hours for JOB Slow, 1.5 hours for TPC-H; $\sim 0.5$ hours faster on \dbms due to its smaller search space)}.
\sys continues to improve and reaches its peak performance after around {4--5 hours}. %
TPC-H has less room for optimization{---it has much fewer joins---}so \sys converges faster.

\emph{\bf Data efficiency.}
\autoref{fig:sample-curve} shows data efficiency curves. %
{It takes a few thousand executions to reach the experts' performance (on \pg: 3.2K for JOB, 7.4K for JOB Slow, 0.7K for TPC-H; on \dbms, $\sim 60\%$ fewer plans are needed).}
The number of query plans required is higher for workloads where the agent starts with slower performance.
{Therefore, experiencing more plans helps \sys improve performance by a greater amount.}

{
\textbf{Non-parallel training wall-clock.}
Throughout our evaluation, including the discussions above and \autoref{fig:learning-efficiency}, we configure \sys to use a few query execution nodes per run (average: 2.5 nodes/run) to speed up training. For completeness,
\autoref{fig:wallclock-non-par} shows non-parallel training times where each run uses one execution node.
In all cases, peak performance is reached within single-digit hours, a comfortable ``nightly maintenance'' range.
The time to match the experts is at most 3 hours slower than that for the parallel mode.
}

\noindentparagraph{\bf Sources of speedup.}
\autoref{fig:speedup} shows \sys's per-query speedups over \pg plans.
For JOB, \sys produces better query plans for most queries in both training and testing.
Notably, \sys considerably speeds up the slowest queries.
Slowdowns mostly occur in the queries that are inherently fast to execute, and hence minimally affect the overall runtime.
A similar trend holds for TPC-H.

\noindentparagraph{\bf Summary.}
{
\sys can bootstrap from a minimal cost model and learn to surpass both an open-source and a commercial expert optimizer.
\sys is efficient to train, needing a few hours to match the experts and thousands of plans to reach its peak performance.
}

\begin{figure}[t]
     \centering
     \begin{subfigure}[b]{0.236\textwidth}
         \centering
          \includegraphics[width=\columnwidth]{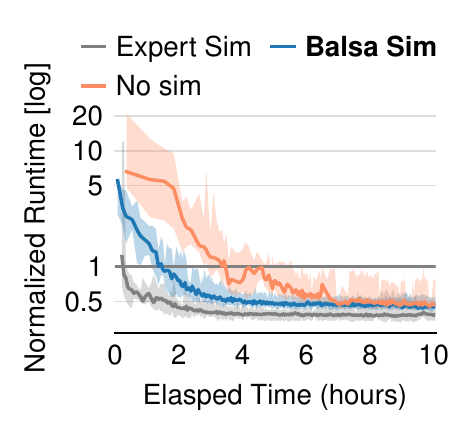}
\vspace{-.23in}
         \caption{Training performance\label{fig:sim-train-perf}}
     \end{subfigure}
     \begin{subfigure}[b]{0.236\textwidth}
         \centering
          \includegraphics[width=\columnwidth]{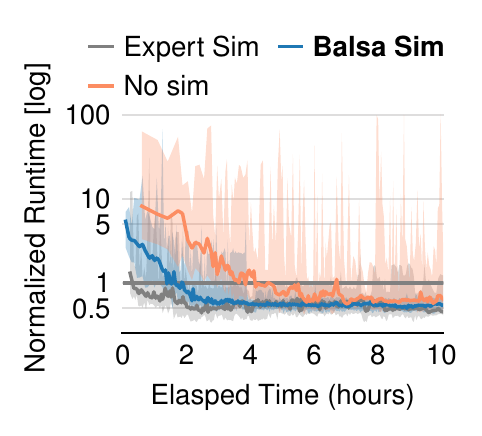}
\vspace{-.23in}
         \caption{Test performance\label{fig:sim-test-perf}}
     \end{subfigure}
\vspace{-.25in}
\caption{{Impact of the initial simulator.
    (a) Better simulators accelerate learning.
    (b) Simulation is essential for generalization.}
\vspace{-.10in}
\label{fig:ablation-sim}}
\end{figure}

\subsection{Analysis of Design Choices}
\label{sec:ablation}
Next, we analyze the design choices of each major component in \sys: (1) the initial simulator, (2) the timeout mechanism, (3) exploration strategies, (4) the training scheme,  and (5) beam search.
In summary, we found all components to positively contribute to \sys's performance and generalization.

In each experiment, we change one component at a time and hold all other configurations fixed at default values.
We then measure each variant's performance on the JOB (random split) workload {on \pg}.
Default choices are highlighted in bold in each figure.

\subsubsection{Impact of the initial simulator.}
\label{sec:ablation-sim}
\sys bootstraps from a minimal simulator.
We can consider two alternatives that differ the most from this choice in terms of the amount of prior knowledge:
\begin{itemize}[leftmargin=*]
\item \textbf{Expert Simulator}: the cost model from {an expert optimizer}, \pg, which has sophisticated modeling of all physical operators and captures the nuances of its execution engine.
  {(Note that this variant means \sys uses this cost model as the simulator; it does not represent \pg's own plans.)}
\item \textbf{\sys Simulator} (\secref{sec:sim_learning}; \cout): a minimal cost model that sums up the estimated result sizes of all operators. It has no knowledge about physical operators or the execution engine.
\item \textbf{No simulator}: skip bootstrapping altogether and initialize the agent from random weights.
\end{itemize}
\autoref{fig:ablation-sim} shows the simulator's impact.
We make four observations:

First, simulators with more prior knowledge
shorten the
time to reach expert performance on training queries (\autoref{fig:sim-train-perf}).
\sys with an expert simulator needs only {$\sim 0.3$ hours} of learning to match the expert.
\sys's default simple simulator takes {$\sim 1.4$ hours} to match, while agents without simulation learning take {$\sim 3.8$ hours}.

Second, more prior knowledge also leads to slightly better final performance at the end of training (\autoref{fig:sim-train-perf}).
The gap, however, is relatively small.
Agents using a minimal simulator mostly catch up with those using an expert simulator. %

Third, it is a pleasant surprise that the  agents without simulation (``No sim'') can finish training.
This is enabled by the use of  timeouts and safe exploration, which keep the bulk of the learning safe.

Fourth, {\emph{simulation is essential for generalization}.}
Agents without simulation learning can fail at test time ({note the high variance of ``No sim'' in \autoref{fig:sim-test-perf}}).
The unstable performance on test queries occurs despite good training performance, rendering this choice impractical.
The instability is caused by randomly initialized agents overfitting the experience collected during the real execution phase, which is limited in quantity ($\sim 700$ subplans per iteration, so it takes at least $\sim 700$ iterations to catch up to the 0.5M-plan simulation dataset, assuming each iteration's data is unique).

In summary,
bootstrapping from a minimal simulator gives good train and test time performance.
Since new execution engines may not  have an expert-developed cost model,
this approach has the additional benefit of potentially generalizing to new systems and alleviating the human development cost.

\begin{figure}[t]
     \centering
     \begin{subfigure}[b]{0.236\textwidth}
         \centering
          \includegraphics[width=\columnwidth]{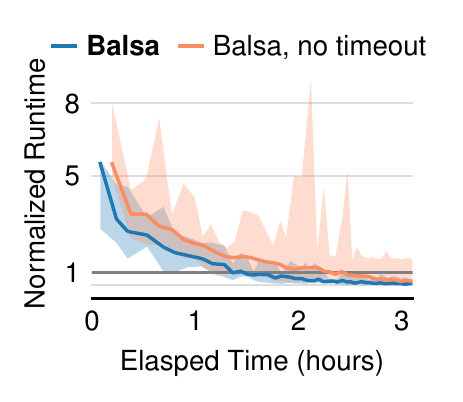}
\vspace{-.23in}
         \caption{Initial training performance\label{fig:timeout-train-perf}}
     \end{subfigure}\hfill
     \begin{subfigure}[b]{0.236\textwidth}
         \centering
          \includegraphics[width=\columnwidth]{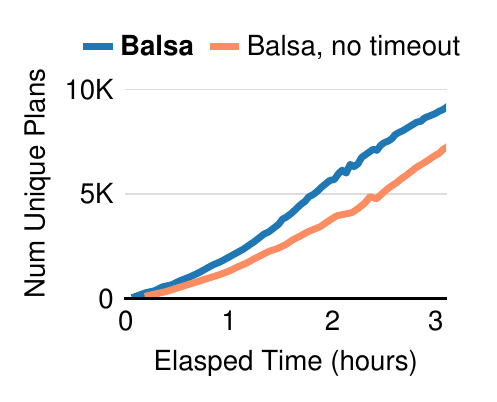}
\vspace{-.23in}
         \caption{Number of executed plans\label{fig:timeout-num-execs}}
     \end{subfigure}
\vspace{-.1in}
\caption{{Impact of the timeout mechanism.}
  (a) Timeouts accelerate learning and prevent spikes.
  (b) With the same wall-clock time, agents with timeouts execute more plans, improving plan variety.
\vspace{-.10in}
\label{fig:ablation-timeout}}
\end{figure}

\subsubsection{Impact of the timeout mechanism.}
We study the impact of timeouts (\secref{sec:timeout}), a mechanism critical for real execution learning:
\begin{itemize}[leftmargin=*]
\item \textbf{Timeout}: early-terminate query plans that have been executing for longer than the current iteration's timeout.
\item \textbf{No timeout}: the mechanism is turned off.
\end{itemize}
With timeouts, agents are expected to save wall-clock time on unpromising plans and potentially learn faster.

Results are presented in \autoref{fig:ablation-timeout}.
Timeout agents reach expert performance about
{35\%}
faster than no-timeout agents (\autoref{fig:timeout-train-perf}).
While both choices lead to similar final performance, there is a pronounced difference in the initial phase of learning.
Agents without timeouts may execute expensive query plans, leading to significant spikes.
Such regressions are unpredictable: they can happen after the no-timeout agents reaching expert performance.

In contrast, agents achieve safe execution when timeout is enabled.
The early-terminated plans ``nudge'' the agents in a different direction to look for more promising plans.
\autoref{fig:timeout-num-execs} shows how the saved time is more judiciously spent:
with the same wall-clock time, agents with timeouts run more plans, speeding up learning.

Overall, these results show that the timeout mechanism accelerates learning and improves \sys's plan variety.

\subsubsection{Impact of exploration.}
\label{sec:ablation-explore}
Exploration exposes RL agents to diverse states, boosting performance and generalization.
We compare:
\begin{itemize}[leftmargin=*]
    \item \textbf{Count-based exploration} (\secref{sec:exploration}): \sys's safe exploration method, which chooses the best unseen plan from beam search outputs.
    \item \textbf{$\epsilon$-greedy beam search}: at each step of the search, with a small probability $\epsilon$ the beam is ``collapsed'' into one state, discarding the rest. The search continues as usual. We chose
      $\epsilon$
      such that about 10\% of training queries have random joins injected.
    \item \textbf{No exploration}: no exploration algorithms are used.
\end{itemize}

\begin{figure}[t]
     \centering
     \begin{subfigure}[b]{0.236\textwidth}
         \centering
          \includegraphics[width=\columnwidth]{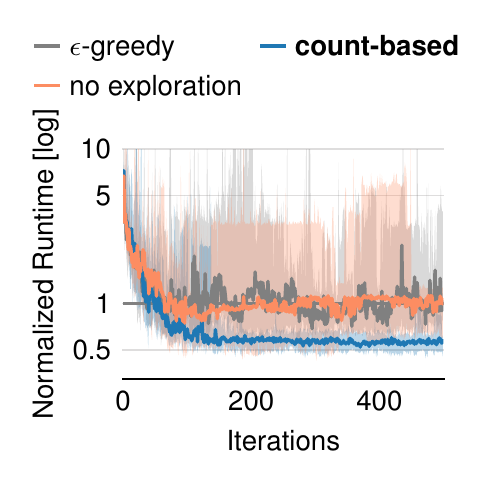}
\vspace{-.23in}
         \caption{Test performance\label{fig:explore-learning}}
     \end{subfigure}\hfill
     \begin{subfigure}[b]{0.236\textwidth}
         \centering
          \includegraphics[width=\columnwidth]{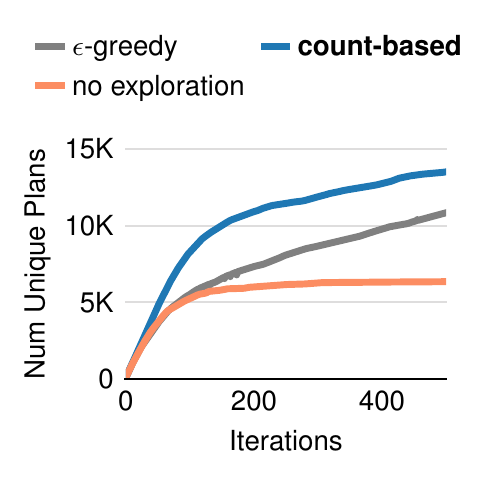}
\vspace{-.23in}
         \caption{Number of unique plans seen\label{fig:explore-num-execs}}
     \end{subfigure}
\vspace{-.1in}
\caption{{Impact of exploration.
  \sys's count-based safe exploration improves generalization to unseen test queries.}
\label{fig:ablation-explore}}
\vspace{-.2in}
\end{figure}

{\autoref{fig:explore-learning} shows that agents with count-based safe exploration generalize to test queries much better than the other two variants.}
The better generalization is a result of the higher number of distinct plans experienced (\autoref{fig:explore-num-execs}).
Training performance is omitted for space reasons, where count-based is around {8\% and 14\%} faster than no-exploration and $\epsilon$-greedy beam at convergence, respectively.

Interestingly, although $\epsilon$-greedy beam search has {similar} plan diversity to count-based, it is less stable.
This is because it contains random joins, which may only lead to low-quality complete plans even when a value network is used to guide the remaining search.

{In summary, these results show that safe exploration is non-trivial, and \sys's count-based method is both simple and effective.}

\begin{figure}[t]
     \centering
     \begin{subfigure}[b]{0.236\textwidth}
         \centering
          \includegraphics[width=\columnwidth]{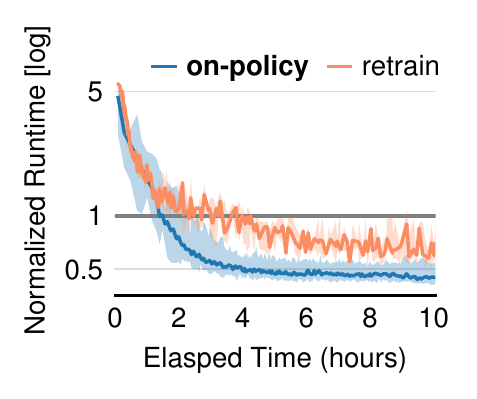}
\vspace{-.23in}
         \caption{Training performance\label{fig:on-pol-learning}}
     \end{subfigure}\hfill
     \begin{subfigure}[b]{0.236\textwidth}
         \centering
          \includegraphics[width=\columnwidth]{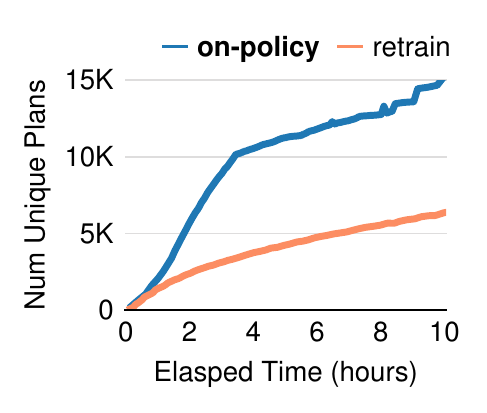}
\vspace{-.23in}
         \caption{Number of unique plans seen\label{fig:on-pol-execs}}
     \end{subfigure}
\vspace{-.1in}
\caption{{Impact of the training scheme.}
  (a) On-policy learning accelerates training. %
  (b) Time saved is used towards more exploration.
\label{fig:ablation-train-scheme}}
\vspace{-.2in}
\end{figure}

\subsubsection{Impact of the training scheme.}
\label{sec:eval:ablation:on-pol}
We compare \sys's on-policy learning  to a full retrain scheme used by prior work, Neo~\cite{neo}:
\begin{itemize}[leftmargin=*]
    \item \textbf{On-policy learning} (\secref{sec:rl-algo}): \sys's training scheme which uses the latest iteration's data to update \vreal.
    \item \textbf{Retrain}: re-initialize  \vreal and retrain on the entire experience ($\mathcal{D}_\text{real}$)  at every iteration. Last iteration's \vreal is discarded.
\end{itemize}

{On-policy learning significantly accelerates training, reaching the expert's performance 2.1$\times$ faster than retrain agents (\autoref{fig:on-pol-learning}).}
Its lead is consistent throughout training.
The faster learning is due to on-policy saving time by updating \vreal on a constant-size dataset, rather than retraining it on an increasingly larger dataset.
The time saved is used towards exploration, i.e., executing more unique plans (\autoref{fig:on-pol-execs}).
Better exploration thus further accelerates learning.
On-policy has slightly higher variance due to performing SGD on much less data.
However, the slowest on-policy agent (the upper edge of the shading) is still mostly  faster than retrain agents.

\begin{figure}[t]
     \centering
     \begin{subfigure}[b]{0.236\textwidth}
         \centering
          \includegraphics[width=\columnwidth]{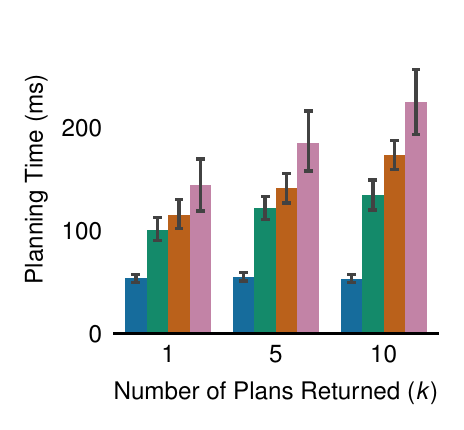}
\vspace{-.23in}
         \caption{Per-query planning time\label{fig:timing-planning-time}}
     \end{subfigure}
     \begin{subfigure}[b]{0.236\textwidth}
         \centering
          \includegraphics[width=\columnwidth]{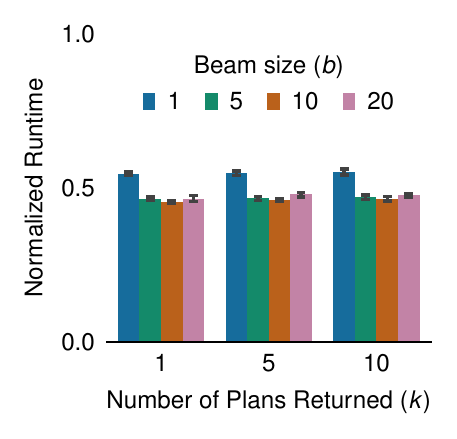}
\vspace{-.23in}
         \caption{Workload runtime\label{fig:timing-runtime}}
     \end{subfigure}
\vspace{-.25in}
\caption{Impact of search parameters on planning time and performance on JOB test set.
Means and standard deviations are shown.
\vspace{-.25in}
\label{fig:timing}}
\end{figure}

\subsubsection{Impact of planning time.}
\sys performs beam search with beam size $b$ {using} the value network to generate $k$ {complete} query plans, {and} then picks the best plan to execute (during training, the best unexplored plan is picked).
\autoref{fig:timing} studies \sys's planning time and performance of the JOB test queries using various combinations of $b$ and $k$ on a trained checkpoint.

For all settings, the mean per-query planning time is below 250ms.
The planner is implemented in Python and thus leaves room for optimization.
Using $b=1$ (where beam search degenerates into greedy search) slightly hurts performance; all other settings produce plans with similar runtime.
Hence, \sys's performance is insensitive to these parameters, and we can flexibly reduce planning time for deployment by using lower values (e.g., $b=5, k=1$ speeds up planning time by $2\times$ with no performance drop).
We use $b=20, k=10$ during training as larger values can help exploration.

\subsection{Comparison with Learning from Expert Demonstrations}
\label{sec:compare-to-neo}

We compare \sys with Neo~\cite{neo}, a recently proposed learned optimizer that relies on \pg-generated plans---i.e., learning from expert demonstrations.
{This experiment uses the same setup as \secref{sec:ablation} (JOB workload on \pg).}
As Neo is not open source, we implement our best-effort reproduction, denoted as ``Neo-impl''.
We make both approaches use identical modeling choices (e.g., architecture, featurizations, beam search), and turn off \sys's algorithmic components for Neo-impl (bootstrapping from simulation; on-policy learning; exploration; timeout mechanism).
One notable difference is that Neo completely resets its model to random weights in each iteration and retrains it on the entire collected experience.

\begin{figure}[t]
     \centering
     \begin{subfigure}[b]{0.236\textwidth}
         \centering
          \includegraphics[width=\columnwidth]{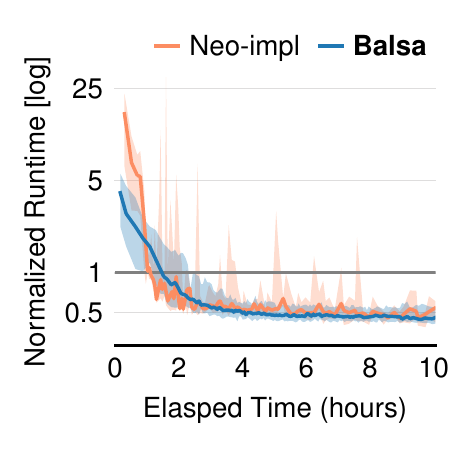}
\vspace{-.23in}
         \caption{Training performance\label{fig:neo-impl-training}}
     \end{subfigure}
     \begin{subfigure}[b]{0.236\textwidth}
         \centering
          \includegraphics[width=\columnwidth]{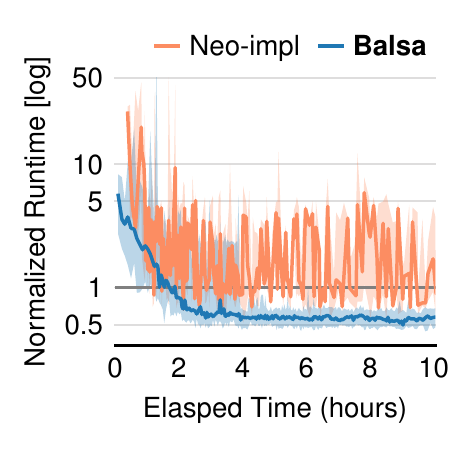}
\vspace{-.23in}
         \caption{Test performance\label{fig:neo-impl-test}}
     \end{subfigure}
\vspace{-.25in}
\caption{{Comparison with learning from expert demonstrations.}
  \label{fig:compare-to-neo}}
\vspace{-.1in}
\end{figure}

\autoref{fig:neo-impl-training} shows training performance.
At initialization, \sys is 5$\times$ faster than Neo-impl, since simulation learning provides a high state coverage (\autoref{table:sim}) as opposed to
 a limited number of
 expert demonstrations (one complete plan per query).
\sys remains stable throughout training, as it employs timeouts.
Neo-impl experiences {performance spikes (note the variance)} as it has no mechanism to deal with disastrous plans.
These regressions are unpredictable and can occur {after hours of training}.
In terms of training efficiency, Neo-impl's retraining scheme makes it progress increasingly slower as the amount of experience accumulates.
Neo-impl spent about {25 hours} to finish 100 iterations, whereas \sys only spent {2.6} hours.

Surprisingly, despite reaching a relatively stable training performance {with 5 hours of learning},
Neo-impl is still not robust enough to generalize to unseen test queries and suffers
from high variance (\autoref{fig:neo-impl-test}).
Its median workload runtime fluctuates between 1--5$\times$ slower than the expert and its maximum is up to {$10\times$} worse.
This failure mode may prohibit this approach from producing reliable models for practical deployment.

{In contrast, \sys is much more robust.}
\sys consistently generates  faster plans  than the expert for unseen queries, with a  2$\times$ maximum speedup.
\sys's better generalization is due to a broader state coverage offered by simulation, on-policy learning, and safe exploration (see Figures~\ref{fig:ablation-explore} and~\ref{fig:ablation-train-scheme}).

In sum, \sys learns faster, achieves safe execution, generalizes better due to simulation and better exploration,
{while refuting the previously held belief that expert demonstrations are needed~\cite{neo}.}

{
\subsubsection{Comparison with Bao}
\label{sec:compare-to-bao}

Bao~\cite{marcus2020bao} is a related approach that assumes an expert optimizer is available.
Like Neo, it requires expert demonstrations to train its model.
Bao learns to provide a set of hints (e.g., disable hash join) for each query, ``steering'' the expert optimizer to produce better plans.
This is different from \sys which learns to produce physical plans by itself.
Nevertheless, we compare the performance of the query plans generated by \sys with those by Bao on top of \pg.

We substantially optimize the Bao source code~\cite{bao_code} as follows.
First, we turn on an optimization that bootstraps its model from \pg's expert plans, rather than from a random state.
Second, its paper specifies that it trains on the most recent $k=2000$ experiences, which we found led to highly unstable performance. We thus train Bao on all past experiences, stabilizing convergence.

\vspace{-.05in}
\begin{table}[h]\centering\small%
  \caption{{\sys vs. Bao: speedups with respect to \pg.}\label{table:bao}}
\vspace{-.15in}
\begin{tabular}{@{} l r r r r @{}} \toprule
{} & {JOB, train} & {JOB, test} & {JOB Slow, train} & {JOB Slow, test} \\   \midrule
\sys & $2.1\times$ & $1.7\times$ & $1.3\times$ & $1.3\times$ \\
Bao  & $1.6\times$ & $1.8\times$ & $1.2\times$ & $1.1\times$ \\
\bottomrule
\end{tabular}
\vspace{-.05in}
\end{table}
\autoref{table:bao} shows that Balsa generally matches or outperforms Bao.
These results are not surprising: they confirm the finding in the Bao paper that a learned optimizer with \emph{higher degrees of freedom} (action space) can outperform Bao in plan quality on stable workloads.
}

\begin{figure}[tp]
\centering
\includegraphics[width=\columnwidth]{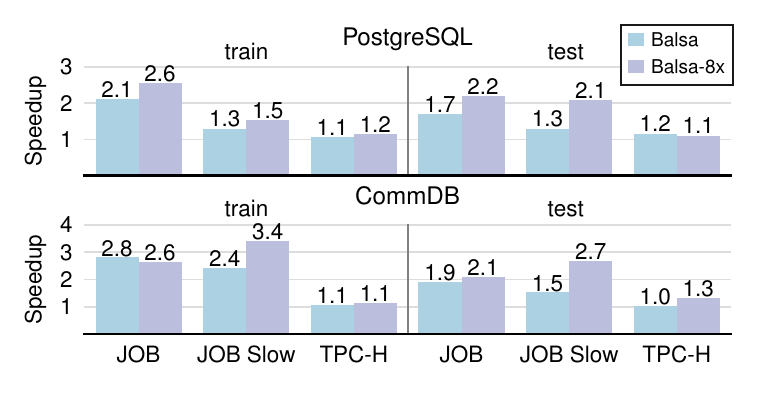}
\vspace{-.35in}
\caption{{Enhancing generalization using diversified experiences.}
  \label{fig:generalization}}
\vspace{-.10in}
\end{figure}

\subsection{Enhancing Generalization}
\label{sec:eval-generalization}
{\autoref{fig:perf} already shows that \sys can generalize to unseen test queries quite well, outperforming experts without ever seeing the test queries.
  Here, we study \emph{(i)} the benefit of diversified experiences (\secref{sec:generalization}), and \emph{(ii)}
  generalizing to entirely distinct join templates/filters.

\noindentparagraph{\bf Diversified experiences.}
We build diversified experiences for all workloads/engines in \autoref{fig:perf}, by merging the data of each main experiment's eight agents.
We retrain a new agent on top,
referred to as ``\sys-8x''; this process is repeated eight times to control for training variance.
(Training is efficient as no query executions are performed.)
\autoref{fig:generalization} shows the median performance: we observe that
\sys-8x \emph{improves speedups on both training and test queries} in almost all cases, sometimes even by 60--80\% (JOB Slow, test).

Improving training speedups is not surprising:
a retrained agent
can mix-and-match the best plans found by the base agents.
Importantly, \emph{test queries see large speedups too without ever being executed} (e.g., on both engines, both JOB splits now have $>2\times$ test speedups).
This is because diversified experiences have highly diverse plans, so more generalizable value networks can be trained on top.}

\noindentparagraph{\bf {Queries with entirely new join templates.}}
{
  We further examine \sys's generalization to difficult unseen queries.
First, we split JOB using 4 slowest \emph{templates} (17, 16, 6, 19) as the test set (20 queries) and the rest as the train set.
On this new split, \sys achieves good train and test speedups (1.4$\times$, 1.5$\times$), further confirming its robustness.}

{Second,} we evaluate on Extended JOB (Ext-JOB), a hard generalization workload~\cite{neo}.
It has 24 new queries on the same IMDb dataset, having 2--10 joins and averaging 5 joins per query.
These queries are challenging and ``out-of-distribution'' since they contain {\emph{entirely different join templates and predicates}} from the original JOB.

\begin{figure}[t]
     \centering
     \begin{subfigure}[b]{0.236\textwidth}
         \centering
          \includegraphics[width=\columnwidth]{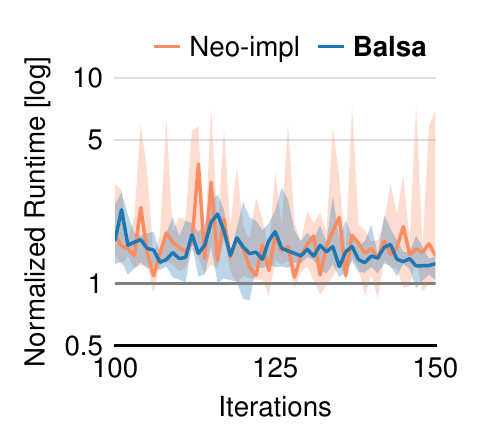}
\vspace{-.23in}
\caption{Single-agent experiences\label{fig:compare-neo-balsa-extjob}}
     \end{subfigure}
     \begin{subfigure}[b]{0.236\textwidth}
         \centering
          \includegraphics[width=\columnwidth]{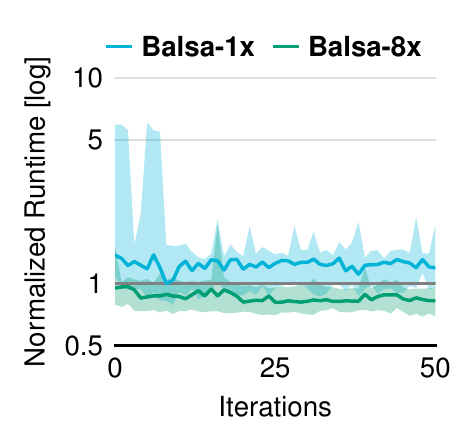}
\vspace{-.23in}
\caption{Diversified experiences\label{fig:compare-reborn-extjob}}
     \end{subfigure}
\vspace{-.25in}
\caption{%
  {Generalizing to highly distinct join templates:
  test performance on Ext-JOB, with JOB as the training set. On \pg}.
\label{fig:compare-offline-ext-job}}
\vspace{-.15in}
\end{figure}

Figure~\ref{fig:compare-neo-balsa-extjob} shows the test performance of Neo-Impl and \sys on Ext-JOB with the entire 113 JOB queries as the training set.
While \sys is  more stable than  Neo-impl, {neither surpasses the expert on the Ext-JOB test set (although they come close)}. %
This confirms that Ext-JOB is a highly challenging generalization workload.

{Next, we compare \sys-8x as described above, with \sys-1x that retrains on only one agent's data.
Surprisingly, in iteration 0, \sys-8x already matches the expert on the test set (\autoref{fig:compare-reborn-extjob}).
We then allow these agents to learn for 50 more iterations on the training set.
Throughout the  process, \emph{the agents never train on the Ext-JOB test queries}.
\sys-8x reaches significantly better test set performance on Ext-JOB (20\% faster than the expert) than \sys-1x
(which still fails to match the expert).
The gain is also consistent.
These results show that \emph{diversified experiences}
and \emph{further exploration} are valuable strategies to improve  generalization to out-of-distribution queries.}

\begin{figure}[tp]
    \centering
    \includegraphics[width=\columnwidth]{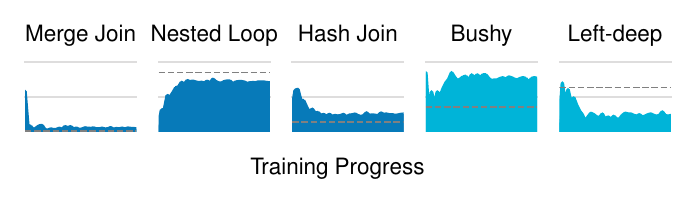}
\vspace{-.40in}
\caption{%
  {\sys's use of join operators (dark blue) and plan shapes (light blue) on JOB. Dashed lines are counts from \pg plans.}
    \label{fig:behavior}}
\vspace{-.15in}
\end{figure}

\subsection{Behaviors Learned by \sys}
\label{sec:behavior}
To gain intuition on the behaviors learned by \sys, we visualize the operator and shape compositions of agent-produced plans over the course of training.
Results are shown in \autoref{fig:behavior}.

In early stages of training,
\sys quickly learns to reduce the use of operators and shapes that incur high runtimes in the current environment. %
For example, after 25 iterations, the use of merge joins is kept below 10\%. %
Meanwhile, \sys starts to prefer more efficient choices.
Nested loop joins are preferred since a large portion (85\% across iterations) are the efficient indexed variant.

\sys's preference is novel when compared to the expert, a difference especially pronounced in the plan shapes.
This is due to the expert optimizer being one-size-fits-all, while \sys learns to tailor to the given workload and hardware.

\section{Related Work}
\label{sec:related}

\noindentparagraph{\bf Learned query optimizers.}
\sys is most related to DQ~\cite{dq} and Neo~\cite{neo}.
DQ offers the insight that the classical components of query optimization---cost estimation and plan enumeration---can be cast as long-term value estimation and planning.
All three work follow this formulation by using a learned value network and plan search. %
\sys also adopts DQ's use of batched data collection on top of a cost model in our simulation learning.
Unlike DQ, \sys demonstrates fine-tuning entire workloads in real execution.

Neo requires learning from expert demonstrations (\pg plans) followed by fine-tuning.
{In contrast, \sys does \emph{not} learn from an expert optimizer.
Lifting this restrictive assumption opens the possibility to automatically learn to optimize in future environments.}
\sys differs in three more aspects with important consequences.
\emph{(i)} Learning from a simulator fundamentally differs from expert demonstrations.
{While the latter are inherently limited in quantity and variety (one expert plan per query), simulation allows us to extract a maximal amount of experience, boosting generalization.}
\emph{(ii)} \sys addresses the challenge of disastrous and slow plans.
\emph{(iii)} {\sys introduces novel techniques (e.g., on-policy learning, timeout as a learning curriculum, safe exploration, diversified experiences),
  all of which lead to higher efficiency, performance, or robustness.}
In \secref{sec:compare-to-neo}, we showed that \sys outperforms the approach of learning from expert demonstrations and is more robust on unseen queries,
despite not {learning from} an expert optimizer.

SkinnerDB~\cite{skinnerdb_sigmod} is {an execution algorithm that learns by trying many left-deep join orders during a query's execution.}
{Both \sys and SkinnerDB use timeouts to mitigate bad plans but propose substantially different timeout policies.
  While SkinnerDB must iterate over a set of pre-defined timeouts unrelated to prior executions,
  \sys directly uses past plans' latencies as timeouts.
  \sys also offers more general capabilities, as it can build bushy plans and assign physical operators, both of which are not supported in SkinnerDB.}

\noindentparagraph{\bf {Optimizer assistants.}}
{%
Many recent proposals use ML to \emph{assist or improve existing optimizers}.
Since Leis \etal~\cite{leis2015good} showed that inaccurate cardinality estimates are most responsible for poor plans,
many projects have used ML to improve cardinality estimation~\cite{wang2020we,msrlearnedcard,kipf2018learned,msr-lightweight,deepdb,naru,neurocard,shetiya2020astrid,e2e_cost}, thus helping today's optimizers find better plans. %
The recent work Bao~\cite{marcus2020bao} also assists expert optimizers by learning what optimizer flags to set for each query.
Different from this line of work, \sys does not assist an existing optimizer, and tackles learning to optimize precisely assuming no expert optimizers.}

\noindentparagraph{\bf {Sim-to-real, timeouts, and caching}}
{
are general techniques applicable to a range of systems problems.
Hilprecht \etal~\cite{tum_part_advisor} have proposed using sim-to-real to learn high-quality data partitionings and applying timeouts and caching to optimize training.
\sys applies these methods in learned query optimization instead and offers the novel finding that simulation learning improves generalization.}
\section{Lessons Learned and Discussions}
\label{sec:discussions}

During the development of \sys, we have learned a few lessons. We discuss them below.

\noindentparagraph{\bf {Simulation learning boosts generalization.}}
{%
  To our surprise,
  while \sys generalizes well to unseen queries,
  we find that agents without a simulation phase---including those that learn from expert demonstrations---become unstable on new queries (\secref{sec:ablation-sim}, \secref{sec:compare-to-neo}).
  At first glance, it might be counterintuitive why simulation improves generalization.
  After all, the simulator we use is a minimal, logical-only cost model that is agnostic to the execution environment.}
  It imparts inaccurate knowledge to the agent that must be corrected.

  We believe the reason is the simulation enables \sys to achieve \emph{a high coverage of the plan space}.
During bootstrapping, \sys trains on thousands of plans per query (\autoref{table:sim}), much more than the experiences collected in real execution.
Then, in real execution, a bootstrapped agent {can update its belief to simultaneously correct much of the simulated knowledge}, which can improve generalization.
  In contrast, agents that learn only from real executions will only see a small set of query plans, which can lead to overfitting.

\noindentparagraph{\bf {Using inaccurate cardinality estimates.}}
{
In traditional optimizers, cardinality estimates are known to be highly inaccurate~\cite{leis2015good}, which can lead to poor plans.
In \sys, however, we find an effective use of inaccurate estimates: use them in the simulator.
We find that inaccurate estimates can still provide effective simulation\footnote{{We use \pg's estimates, which have $\sim 100\times$ median errors and up to $10^6\times$ tail errors on JOB~\cite{leis2015good}. We tried making them even more inaccurate, by dividing them by \emph{random noises} (a median noise factor of 5$\times$), and saw little impact on \sys's plans.}}.
Importantly, \sys's performance is not overly tied to the simulator---most learning occurs \emph{after} simulation,
when \sys uses real execution to vastly improve over the simulated knowledge (e.g., initial vs. final performance have a 4--40$\times$ gap in \autoref{fig:learning-efficiency}).
Consistent with prior work~\cite{leis2015good}, we expect better estimates to lead to a better simulator, which would accelerate learning (e.g., ``Expert Sim'' in \autoref{fig:ablation-sim}).
}

\noindentparagraph{\bf {How to better leverage an expert optimizer, if available?}}
{
  For learning to optimize in a new system, even if a \emph{compatible} expert optimizer (i.e., all operators of the expert are supported by the target engine) exists,
  prior state-of-the-art~\cite{neo} proposes bootstrapping only from the expert optimizer's \emph{plans}.
  We show that this can lead to poor generalization due to the limited amount of demonstrations (\secref{sec:compare-to-neo}).
  In contrast, \sys can \emph{better leverage the expert} by bootstrapping from the expert optimizer's \emph{cost model}---a data-rich simulator (see the ``Expert Sim'' \sys variant in \autoref{fig:ablation-sim}).
  We show that bootstrapping from a cost model significantly improves generalization to new queries (\secref{sec:ablation-sim}), which is a novel finding of this paper.

}

\section{Conclusion}
\label{sec:conclusion}

To our knowledge, \sys is the first approach to show that %
{learning an optimizer without expert demonstrations}
is both possible and efficient.
\sys learns by iteratively
planning
a given set of queries, executing them, and learning from their latencies to build better execution plans in the future.
{To make learning practical}, \sys must avoid
{disastrous}
plans that can dramatically
{hinder}
learning.
We address this key challenge with three simple techniques:
bootstrapping from a simulator,
safe execution, and safe exploration.

\sys paves the road towards
automatically learning a query optimizer tailored to a workload and a compute environment.
New data systems may have execution models~\cite{modin} or objectives~\cite{materialize_blog} that go beyond  our knowledge of query optimization.
By learning on its own and
{not learning from}
an expert system, \sys may alleviate the significant optimizer development cost for systems yet to be developed.
\sys is a first step towards this exciting direction.

\bibliographystyle{ACM-Reference-Format}
\balance
\bibliography{refs}

\end{document}